\newcommand{\CLASSINPUTinnersidemargin}{0.5in}
\newcommand{\CLASSINPUToutersidemargin}{0.5in}
\documentclass[10pt,journal,compsoc, letter]{IEEEtran}
\ifCLASSOPTIONcompsoc
  \usepackage[nocompress]{cite}
\else
  \usepackage{cite}
\fi
\ifCLASSINFOpdf
\else
\fi

\usepackage{graphicx} 
\usepackage{subfig}
\usepackage{balance} 
\usepackage{xcolor}

\begin{document}
%
\title{To Explore What Isn't There --- Glyph-based Visualization for Analysis of Missing Values}
%
%
%
%

\author{Sara~Johansson~Fernstad
        and~Jimmy~Johansson
\IEEEcompsocitemizethanks{\IEEEcompsocthanksitem  Sara Johansson Fernstad is with the School of Computing, Newcastle University, UK.\protect\\
E-mail: sara.fernstad@newcastle.ac.uk
\IEEEcompsocthanksitem Jimmy Johansson is with Link\"{o}ping University, Sweden.}
}
\IEEEtitleabstractindextext{%
\begin{abstract}


This paper contributes a novel visualization method, Missingness Glyph, for analysis and exploration of missing values in data. Missing values are a common challenge in most data generating domains and may cause a range of analysis issues. Missingness in data may indicate potential problems in data collection and pre-processing, or highlight important data characteristics. While the development and improvement of statistical methods for dealing with missing data is a research area in its own right, mainly focussing on replacing missing values with estimated values, considerably less focus has been put on visualization of missing values. Nonetheless, visualization and explorative analysis has great potential to support understanding of missingness in data, and to enable gaining of novel insights into patterns of missingness in a way that statistical methods are unable to. The Missingness Glyph supports identification of relevant missingness patterns in data, and is evaluated and compared to two other visualization methods in context of the missingness patterns. The results are promising and confirms that the Missingness Glyph in several cases perform better than the alternative visualization methods.
\end{abstract}

\begin{IEEEkeywords}
Missing data, information visualization, glyphs.
\end{IEEEkeywords}}

\maketitle

\IEEEdisplaynontitleabstractindextext

%
\IEEEpeerreviewmaketitle

\IEEEraisesectionheading{\section{Introduction}\label{sec:introduction}}

%
%
%
%

\IEEEPARstart{D}{ata} sets with missing values, commonly known as incomplete or missing data, are a frequent challenge in data analysis across a range of domains. They are known to cause issues such as biased, uncertain and unreliable results. A large number of statistical methods have been developed for dealing with missing values, which are mainly focused on replacing missing data with plausible values (known as imputation) \cite{Carpenter2013}. Meanwhile, the visualization and visual analysis of missing data is a largely overlooked topic, even though visualization has great potential to support understanding and knowledge generation from incomplete data. The awareness of the existence of missing values and the patterns relating to these missing values can be improved by visualization, and through this many potential issues and data uncertainties can be highlighted. In addition to supporting identification of issues arising during data generation and pre-processing, visualization of missing data can reveal important patterns, such as patients missing appointments in medical studies where the missingness may highlight a health issue. Missing data visualization can also facilitate decision making as to how missing values can be most appropriately dealt with. The application of suitable statistical methods for imputation require understanding of the patterns of missingness, and questions such as if records are missing at random or if there are structured patterns needs to be taken into consideration. 

Research by Johansson Fernstad \cite{Fernstad2019} and Johansson Fernstad and Glen \cite{Fernstad2014} suggested a set of missingness patterns of particular importance for analysis. Johansson Fernstad \cite{Fernstad2019} provided context to the complexities of analysing incomplete data, and identified issues to address through evaluation of visualization methods. This paper presents the Missingness Glyph (MissiG), a novel visualization method for analysis of missing values in multivariate data. MissiG represent both univariate and multivariate patterns of missingness as well as relationships between missing and recorded values, to support the identification and understanding of missingness patterns, and through this highlight uncertainty to aid decision making. To ensure its usability, MissiG is designed based on well established glyph design principles \cite{Chung2015, Maguire2012, borgo2013}, utilizing simple and clearly separable visual channels (colour, height and shape) for category separation, magnitude representation and distribution comparison. It is designed to work equally well as a standalone visualization or as an enhancement to existing multivariate data visualization methods. To demonstrate its versatility, the paper presents two standalone layout options for MissiG, linear and radial, as well as examples of how it can be used as enhancement for Heatmap and Parallel Coordinates (PC). The usability of MissiG is demonstrated through two evaluations that compare MissiG with a Heatmap, which represents missing values using colour, and with PC where missing values are represented through location. The results indicate that MissiG is generally better or equally good as other visualization methods when it comes to identification of missingness patterns, and that the performance of other visualization methods can be improved by enhancing them with MissiG. The main contributions of this paper are:

\begin{itemize}
\item MissiG, a novel glyph visualization that supports analysis of missingness patterns in data;
\item two layouts (linear and radial) with additional enhancements of MissiG;
\item two evaluations comparing the performance of visualization methods in the context of identification of missingness patterns.
\end{itemize}

\noindent The paper is structured as follows: Section \ref{section::bg} provides background, the MissiG design is described in Section \ref{section:missingnessglyph}. Its usability is demonstrated by examples in Section \ref{section::usecase} and evaluations in Section \ref{section:evaluation}, with conclusions in Section \ref{section:conclusions}.

 

\section{Background}\label{section::bg}
Missing records may occur in any type of data (numerical, categorical, text, relational networks etc.), and the most appropriate method for visualizing missing data will depend on the type of the recorded data. The focus of this paper lies mainly on missing values in multivariate (numerical) data. The visualization presented in the paper can, however, easily be adapted to categorical data and the missingness patterns they are based on are equally applicable to numerical and categorical data. The identification of missing values in data can be a challenging pre-processing step to data analysis, since missing values can be represented in a range of different ways during data collection. While the identification of missing values is an important challenge that can be facilitated by visualization, it is not within the scope of this paper. The contributed visualization method assume that missing values are explicitly marked in the data. This section will provide a brief overview of the analysis of data with missing values. More in depth discussions can be found in, for instance, Johansson Fernstad \cite{Fernstad2019}.

\subsection{Analysis of Missing Data}
The effect missing values have on analysis results depends both on the missingness mechanism, described in \ref{section::misspatterns}, and on how the missing values are handled. It may also be greatly affected by the degree of missingness and distribution of missing values across the data set. The two main approaches to dealing with missing values are removal and imputation. Removal is when data items with missing values are removed prior to analysis. This approach carries a considerable risk of biased results, unless the values are missing completely at random. Imputation is when missing values are replaced by estimated values. There exist a large number of imputation methods, ranging from replacement with arithmetic mean or random draws from representative distributions, to complex multiple imputation methods that combine several imputations following a set of rules \cite{Carpenter2013}. Imputed values may bias and affect the analysis results, depending on the appropriateness of the imputation method.

\subsection{Missingness Patterns} \label{section::misspatterns}
The missingness mechanism \cite{Rubin1976} is a model of how the probability of an observation being missing depends on its own value and on the values of other variables. There are three mechanisms defined: {\it Missing Completely at Random}, {\it Missing at Random} and {\it Missing Not at Random}. The missingness mechanisms are rarely known prior to analysis, they are fairly complex and may be difficult to apply to an exploratory analysis approach. More recent research suggests patterns that may be more straightforward for describing missingness in data. Wang and Wang \cite{Wang2007} suggested three patterns in context of classification data, focussing on the distribution of missing values. In a later paper \cite{Wang2009} they described concepts of relevance for understanding the impact of missing values on the analysis results, addressing both missing values and the relationship between missing and recorded. Based on previous research and interviews with data science practitioners, Johansson Fernstad \cite{Fernstad2019} defined a set of three missingness patterns of relevance for analysing missingness in data, as described below. 

{\bf Amount Missing} (AM) refers to the relative amount of missing values in a variable or a data item, and supports understanding of the distribution of missing values in the data set. Insight into AM in variables can, for example, support identification of variables where the missingness may be particularly difficult to deal with, or highlight subsets of data where conclusions drawn from the recorded values may be unreliable due to the large amount missing values. It can also be useful to investigate whether the missingness may be randomly distributed, since the amount missing would be relatively equal across the data set for random missingness.

{\bf Joint Missingness} (JM) is a multivariate or pairwise pattern that refers to the amount of data items that have missing values in more than one variable at the same time. The pattern may, for example, occur in survey data where participants who refuse to answer a specific question also tend to not answer another specific question. Identification of JM can support discovery of issues in data collection or pre-processing that cause missingness to propagate across the data, as well as identification of data subsets where missingness may need to be dealt with differently to missingness in subsets with less JM.

{\bf Conditional Missingness} (CM) is a pairwise pattern that describes the relationship between items that are missing in one variable and their recorded value in other variables. It aims to describe patterns where the probability of missingness is conditional upon recorded values, and as such supports understanding of relationships between missing and recorded. Investigation of CM can be useful to understand the cause of missingness, and can support decisions on how to deal with the missingness. For example, if missing values in variable $A$ tend to have low recorded values in variable $B$, then imputation of missing values in $A$ based on items with low values in $B$ may be more valid than imputation based on all items. 

As discussed in Johansson Fernstad \cite{Fernstad2019}, these three patterns bring together the main characteristics of the previously suggested missingness patterns. They also provide a more straightforward description of patterns than the missingness mechanism. The research presented in this paper address methods for exploration and identification of missingness in data based on the concepts of AM, JM and CM.

\subsection{Missing Data Visualization}
It can often be meaningful to consider missing values as information bearing signals, rather than issues that need to be removed, since they may provide valuable information and highlight potential issues in data gathering, pre-processing and analysis processes. Fielding et al.\ \cite{Fielding2009} and Djurcilov and Pang \cite{Djurcilov2000} provide examples from health-related surveys and meteorological studies where the absence of data is more informative than an estimated value, and emphasize the value of visualization for understanding of missingness in data.

Shape Coding \cite{Beddow1990} is an early example of representing missing values with colour to support identification of missingness related patterns in multivariate data. Twiddy et al.\ \cite{Twiddy1994} adopted a similar approach where recorded and missing values were visually separated using a colour scheme. MANET \cite{Unwin1996, Theus1997} was another early example where visual representations of missing values were incorporated in the visualization software. The XGobi \cite{Swayne1998} and gGobi \cite{Swayne2003} systems focussed on exploration of missingness, and represented missing data by imputed values which were linked to a separate view to keep track of missing values. Wang and Wang \cite{Wang2007} presented a visualization method for missing values in classification data, utilizing self-organizing maps \cite{Kohonen1998} for clustering, with main focus on whether the missing values were randomly distributed, unevenly distributed or biased towards a particular class. Additionally, a number of R-packages support visualization of missing values, as described by for examples Unwin \cite{unwin2015graphical}. Some of the more recent packages include Naniar \cite{tierney2019naniar}, which includes a range of table and barchart based visualization as well as an UpSet \cite{lex2014upset} style set visualization; and extracat \cite{pilhoefer2014extracat} including the Visna visualization of missing values. VIM (Visualization and Imputation of Missing values) \cite{Templ2012}, utilize various visual attributes to highlight missingness in histograms, scatter plots, PC and other common visual representations. The miP (multiple imputation plots) package \cite{Brix2011} use VIM to visualize imputation results from a range of packages. Cheng et al.\ \cite{Cheng2015} developed an R-package that distinguish imputed missing values from recorded values by colour.
Some R-packages, such as AmeliaView \cite{Honaker2011}, also provide graphical user interfaces for manipulation and control of imputation methods. While interesting, a large part of previous research in missing data visualization may not be able to efficiently deal with growing data sizes. Many methods are focussed on supporting imputation and representing imputed values, which is an important challenge that visualization can facilitate. Such methods may, however, be less appropriate for explorative analysis and knowledge generation. 


Considering missing values as part of the broader issue of data quality, a range of tools have been designed to support data profiling and cleaning. Profiler \cite{kandel2012} utilise inference and data mining approaches to identify quality issues, and use visualization to investigate them in context of the larger dataset. 'Know-your-enemy' \cite{Gschwandtner2018} use a similar approach, with automated quality checks to support visual exploration of quality issues in time series data. Triana et al.\ \cite{Triana2019} utilise data quality dimensions to enrich visualization with quality information. Schulz et al.\ \cite{Schulz2017} defined missing data as one of several data descriptors, and used PC with lines intersecting a point below the axis to represent missing values. Cedilnik and Rheingans \cite{Cedilnik2000} represented uncertainty using procedurally generated annotations, and represented missing data by a distance based probability value. Xie et al. \cite{Xie2006} focussed on data quality and used imputation to obtain quality values for missing data. Arbesser et al.\ \cite{Arbesser2017} presented a system where missing data is one of several quality classes represented by colour. These papers define missingness as one of several quality descriptors in the more general context of data quality, thus not focussing on visual analysis and representation of the specific features of missingness, as is the focus of the work presented here.

Although the importance of visualization of missing data has been emphasized \cite{Wong2012, Kirk2014, Fernstad2014}, little research has investigated how to best represent missing values in data. Eaton et al.\ \cite{Eaton2005} discussed the impact of missing data on the interpretation of visualization, and evaluated three approaches to representing missing values in line graphs. Their results implied that poor indication of missing values has negative effect on interpretation, and suggest that visualization should be enhanced by dedicated visual attributes, annotation or animation to indicate the existence of missing data. They did, however, not suggest what visual attributes may be most appropriate. A later study by Andreasson and Riveiro \cite{Andreasson2014} evaluated the impact on decision making of three techniques for representing missing values visually (emptiness, fuzziness and emptiness plus explanation), and found that while emptiness plus explanation was the most preferred technique and rendered the highest decision confidence it also resulted in higher risk behaviour. Fernandes et al.\ \cite{Fernandes2018} evaluated the impact on decision making of a set of uncertainty visualization on mobile displays, concluding that CDF plots and quantile dotplots can successfully improve decision making despite the limited display space. Song and Szafir \cite{song2018} investigated four categories of missing value representation in line-graphs and bar charts, defined as highlight, downplay, annotation and information removal. They concluded that highlighting of missing values is usually perceived as higher quality, while downplay and information removal is perceived as lower quality. They also found that visualization using highlighting and annotation while preserving the continuity of the recorded data results in the highest perceived data quality and confidence in results.

Recent work by Johansson Fernstad \cite{Fernstad2019} evaluated the performance of three visualization methods in the VIM package \cite{Templ2012} (scatterplot matrix, heatmap and PC), which are enhanced by visual attributes for representation of missing values. The performance was investigated for tasks relating to the identification of the AM, JM and CM patterns. The results indicated that heatmap with missing values represented by colour generally performed best for tasks relating to AM and JM, while PC with missing values represented by lines intersecting a point above the axis, performed better for CM tasks. Conclusions from that study suggest that it is important to:

\begin{enumerate}
\item Include clear frequency representations through size, possibly combined with colour. This generally needs more attention, and in particularly for visualization that lack a natural frequency representation.
\item Include features that connect missing and recorded, as well as missing and missing, across multiple variables. This is important for identification of multivariate patterns, and in particular need more attention in visualization with limited representation of connections.
\item Avoid separation of missing and recorded values in different sets of representations. This is particularly important for CM patterns, but also since missingness analysis would commonly be part of a more extensive analysis where overall data patterns are of interest.
\end{enumerate}

\noindent It was also concluded that location rarely is suitable as sole representation of missing values, due to the intrinsic meaning of location in methods such as PC and scatterplot, and that additional features should be used to emphasize the missingness.

\section{The Missingness Glyph}\label{section:missingnessglyph}
This section will describe MissiG, a novel visualization method designed to support exploration of missing values in data, based on the missingness patterns and results presented by Johansson Fernstad \cite{Fernstad2019}. MissiG is designed to be usable both as a standalone visualization or as a glyph-style enhancement to be added to multivariate visualization methods, thus utilizing the strenghts of common visualization methods while overcoming their limitations in terms of missing data analysis. While the examples provided in this paper mainly focus on numerical data sets, MissiG can easily be adapted to categorical data.

\subsection{Visual Representation of Missingness Patterns}
\begin{figure}[t]%
       \centering
       \subfloat[][The relative amount missing in each variable is represented as a light blue block.]{
       	\includegraphics[width=3.7cm]{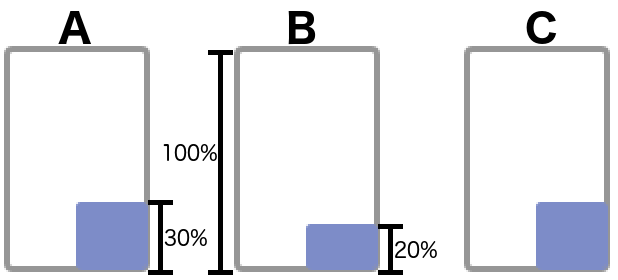}
       	\label{Fig:MissVisA}
       }%
       \qquad
       \subfloat[][Recorded items are represented as histograms, with D being categorical.]{
       	\includegraphics[width=4cm]{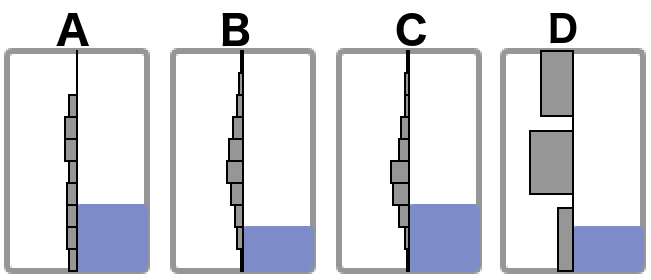}
       	\label{Fig:MissVisB}
       }%
       \qquad
       \subfloat[][The joint missingness with a selected variable is represented as a red block, the selected variable is highlighted in red.]{
       	\includegraphics[width=3.7cm]{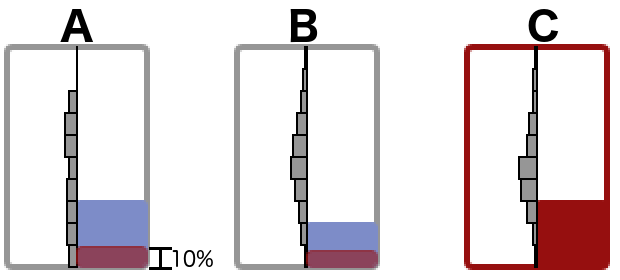}
       	\label{Fig:MissVisC}
       	}%
       \qquad
       \subfloat[][A red histogram is used to represent the distribution of recorded items that are missing in the selected variable.]{
       	\includegraphics[width=3.7cm]{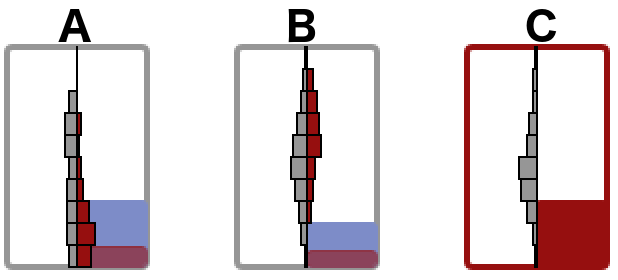}
       	\label{Fig:MissVisD}
       }
       \caption{The basic structure of MissiG for three or four variables. Variable C is selected in \ref{Fig:MissVisC} and \ref{Fig:MissVisD}}%
       \vspace{-2mm}
       \label{Fig:MissVisDescription}%
\end{figure}

In its basic form, each MissiG glyph represent one variable in a data set. The glyph has a rectangular shape using blocks and histograms to represent the data and missingness patterns. As displayed in Fig. \ref{Fig:MissVisA}, the relative amount missing (AM) is represented by the height of a light blue block, where the full length of the main rectangle represents 100\%. In the figure, 30\% of values are missing in the left and right variables, while 20\% of values are missing in the centre variable. With block height being a straightforward representation of frequency, this approach provides an easily interpreted overview of the amount missing in multivariate data sets. The distribution of recorded items in the variable is represented by a grey histogram in the left half of the glyph, as shown in Fig. \ref{Fig:MissVisB}, with low values represented at the bottom and high values at the top. The width of the histogram bins corresponds to the relative number of items with recorded values within the bin range. For a categorical variable, each bin would represent a unique category and the width would correspond to the relative frequency of that category (variable $D$, Fig. \ref{Fig:MissVisB})

The relative number of items that are subsequently missing in a pair of variables (i.e. the JM) and the relationship between missing in one variable and recorded in another (i.e. the CM) both operate on variable pairs and thus there exist a larger number of JM and CM relationships than there are variables. Furthermore, while the JM of a variable pair is non-directional ($A_{JM} \rightarrow B_{JM} = A_{JM} \leftarrow B_{JM}$), the CM is directional ($A_{CM} \rightarrow B_{CM} \neq A_{CM} \leftarrow B_{CM}$). Thus, there are twice as many unique CM relationships as JM relationships in a data set. To avoid visual clutter and increase scalability, MissiG in its basic form only displays representation of JM and CM for a selected variable. A red block is used to represent the JM, as displayed in Fig. \ref{Fig:MissVisC} where variable $C$ is selected and highlighted in red, and where the height of the red block in $A$ indicate that 10\% of items have concurrently missing values in $A$ and $C$. CM is represented through a red histogram in the right half of the glyph, which display the distribution of recorded values in an unselected variable, for the subset of items that have missing values in the selected variable. In Fig. \ref{Fig:MissVisD} variable $C$ is selected, and the red histogram in $A$ represent the distribution in $A$ of items that are missing in $C$ but recorded in $A$. From Fig. \ref{Fig:MissVisD} it is visible that items with missing values in $C$ tend to have relatively low values in $A$ while they have comparably high values in $B$. 

When analysing missingness it is often relevant to understand if the values are missing at random or not. The visual features of MissiG can help out-ruling random missingness through multiple properties. Firstly, where missingness is completely random it can be expected that each variable has roughly the same amount missing; if the height of the blue blocks in the MissiG glyphs varies greatly, we can hence assume that the missingness is not completely random. Secondly, if the missingness is random we can expect certain levels of JM. This expected JM can be defined as $E(\vec{d_j},\vec{d_k})=P(\vec{d_j}) \cdot P(\vec{d_k})$, where $P(\vec{d_j})$ and $P(\vec{d_k})$ are the probabilities that a value is missing in $\vec{d_j}$ and $\vec{d_k}$ respectively. If the JM deviates greatly from $E(\vec{d_j},\vec{d_k})$ we may assume that the missingness is not completely random. Thus, if 50\% of values are missing in variable $A$ and 50\% are missing in $B$, we expect 25\% of values to be concurrently missing in $A$ and $B$ if missingness is completely random. Thirdly, for CM, if the data is randomly missing in $A$, we would expect the overall distribution of recorded values in variable $B$ to be similar to the distribution of recorded values in $B$ for the subset of items that have missing values in $A$. Hence we would expect the grey and red histograms in the glyph for $B$ to have a similar shape. If the shapes of the two histograms differ considerably, we can conclude that the missingness in $A$ is not random and there may be a relationship between recorded values in $B$ and missingness in $A$.

\subsection{Glyph Design Considerations}
Four pieces of information are represented in a glyph: 1) the relative amount missing in a variable; 2) the relative amount jointly missing with another, selected variable; 3) the overall distribution of recorded data; and 4) the distribution of recorded data for items that are missing in another, selected variable. MissiG was designed based on these definitions and a number of well established glyph design principles.

MissiG utilize three main visual channels: colour to distinguish between three categories of information (missing values in a variable, recorded values in a variable, and data relating to another, selected variable), size/height to represent magnitude, and shape for representation and comparison of data distributions. This is based on the principles of Typedness \cite{Chung2015} and Semantic Relevance \cite{Maguire2012}, which emphasize that visual channels should be appropriate for the semanticts of the underlying data, as well as the guidelines suggested by Borgo et al. \cite{borgo2013}. Orderability and Channel Capacity \cite{Chung2015} has also been taken into account, with orderable data (magnitude) being represented by height which has a relatively high capacity, and non-orderable categories being represented by the lower capacity colour channel. Borgo et al. \cite{borgo2013} also emphasize the importance of glyph property interaction normality as well as the use of perceptually uniform properties. Normality and uniformity is maintained in MissiG through magnitude values (AM and JM) being represented relative to the full glyph height, which is the same for all glyphs, and by distribution representations utilizing the full glyph height (this was not completely addressed in an earlier version of the glyph, as described in Section \ref{section:evaluation}).

The importance of simplicity, the use of well known visual channels and well defined rules is highlighted by several design principles (Learnability \cite{Chung2015}, Complexity and Density, Simplicity and Symmetry \cite{borgo2013}). The MissiG design addresses this through the use of basic visual channels representing a clearly defined type of information, with colour representing categories of information (missing values, recorded values, and data relating to a selected variable), height representing magnitudes and shape representing value distributions. The design principles of Separability, Searchability \cite{Chung2015} and Channel Composition \cite{Maguire2012} emphasize the importance of clearly identifiable, non-conflicting visual channels. In the MissiG design this is achieved through the use of visually different and non-overlaying visual channels. Alternative designs that have been considered would generally increase the complexity of the glyph. For example, a pie-chart style representation could have been used for the magnitude (AM and JM), but this would have complicated the additional representation of distributions, as well as loosing the intuitiveness of a minimum and maximum value and, to some extent, the ability to easily compare heights across glyphs. For numerical variables, a violin style representation could have been used instead of histograms, these would however not be viable for categorical variables.

Pop-out effect and saliency of visual channels are important to take into account in glyph design, such that more important information is made more salient. This is, for example, highlighted by the principles of Attention Balance \cite{Chung2015}, Pop-out Effects and Visual Hierarchy \cite{Maguire2012}, and Importance-based mapping \cite{borgo2013}. Colour is one of the most salient visual channels \cite{Quinlan1987}, and more intense colours, such as red, tend to have a stronger pop-out effect. In the MissiG glyph, colour is used to discriminate between the three categories 'missing data', 'recorded data' and 'data related to a selected variable'. Patterns related to a variable that is interactively selected by the user can be expected to be of more interest to the user, thus a more salient colour (red) is chosen to highlight these patterns, while the non-selected missing and recorded values have less salient colours (light blue and light grey). This is also related to the principle of Focus and Context \cite{Chung2015}.

\subsection{Layouts and Enhancements}\label{Sec:layouts}
To provide an as flexible representation as possible, MissiG can be used as a standalone visualization technique with different layouts, or as a glyph-style enhancement to existing methods. In this section two standalone layouts are suggested, with added visual encodings to enhance representation of missingness patterns, but further layouts can also be considered. Furthermore, two examples are provided of how MissiG can be used to extend two common visualization methods.

\begin{figure} [t]%
       \centering
       \subfloat[][AM and distribution of recorded values.]{
       	\includegraphics[width=8.5cm]{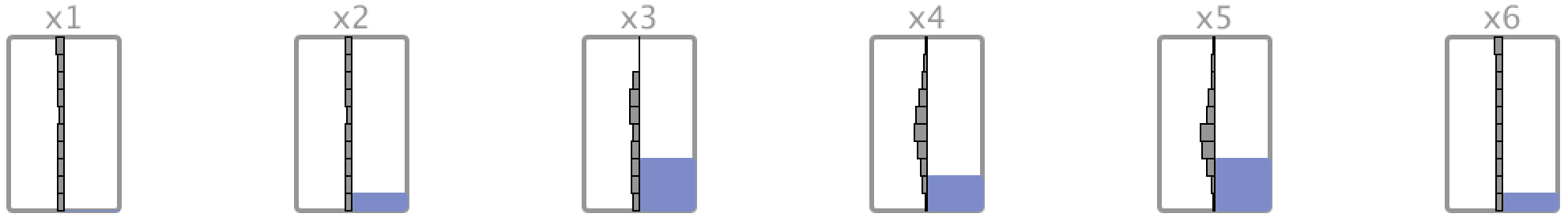}
       	\label{Fig:MissiG_Linear_6Var_noSel}
       }%
       \qquad
       \subfloat[][Variable $x5$ is selected. The highlighted blocks and arcs emphasize JM and red histograms represent CM with missing values in $x5$.]{
       	\includegraphics[width=8.5cm]{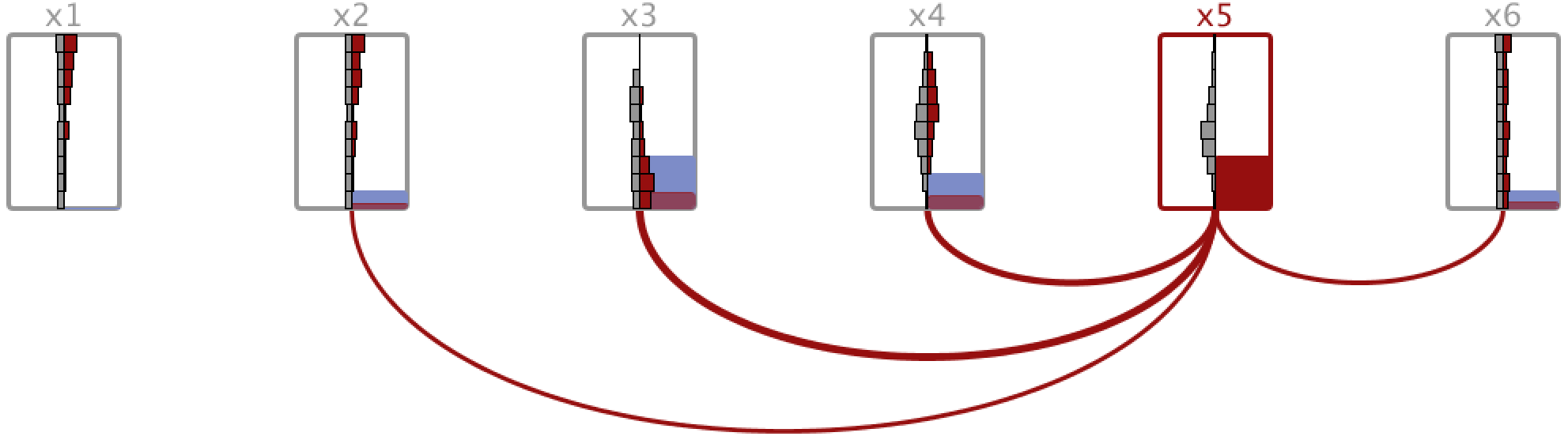}
       	\label{Fig:MissiG_Linear_6Var_x5sel}
       	}%
       \qquad
       \subfloat[][Arcs with thickness corresponding to JM for all variable pairs.]{
       	\includegraphics[width=8.5cm]{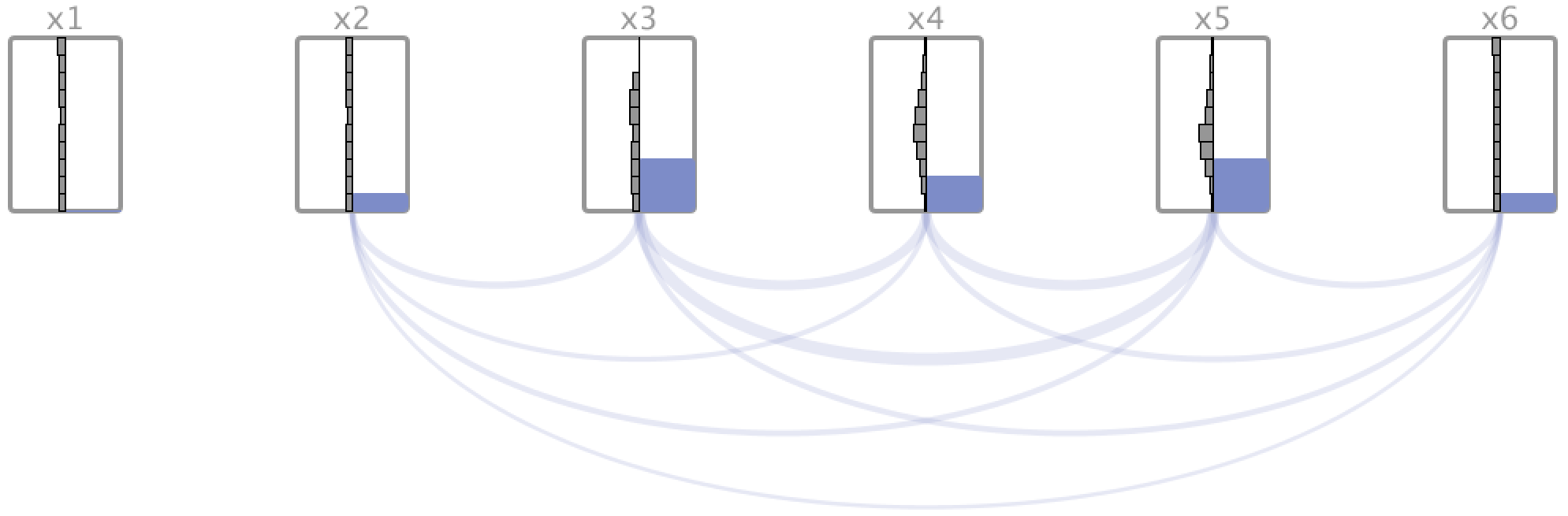}
       	\label{Fig:MissiG_Linear_6Var_allArcs}
       }
       \caption{MissiG with linear layout for a synthetic data set with 6 variables, where $x1$ has no missing values and the remaining variables have 10\% -- 30\% missing.}%
              \vspace{-2mm}
       \label{Fig:MissiG_Linear_6Var}%
\end{figure}

{\bf Linear layout} is the simplest multivariate representation, similar to PC, as displayed in Fig. \ref{Fig:MissiG_Linear_6Var}. This layout facilitates comparison of in particular AM patterns, since the height of blocks are directly comparable across multiple variables. Upon selecting a variable by clicking on it, as displayed in Fig. \ref{Fig:MissiG_Linear_6Var_x5sel} where $x5$ is selected, the JM with the selected variable is enhanced through arcs linking variable pairs. The thickness of the arc represents the JM of the variable pair. Alternatively, the JM of all variable pairs can be represented by arc thickness, as in Fig. \ref{Fig:MissiG_Linear_6Var_allArcs}. In Fig. \ref{Fig:MissiG_Linear_6Var_x5sel} and \ref{Fig:MissiG_Linear_6Var_allArcs} it is, for instance, visible from the thickness of the arcs that $x5$ and $x3$ have relatively high JM, while the JM of $x5$ and $x2$ is lower. The red histograms in the figure also reveal some CM patterns. For instance, the red histograms in the $x1$, $x2$ and $x4$ glyphs, which are denser for higher values than corresponding grey histogram, indicate a relationship between missing values in $x5$ and high values in $x1$, $x2$ and $x4$; while there appear to be a relationship between missing in $x5$ and low values in $x3$. For $x6$, the red and grey histogram have a similar shape, indicating that there is no relationship between missing in $x5$ and recorded in $x6$.

\begin{figure} [t]%
\centering
       	\includegraphics[width=6cm]{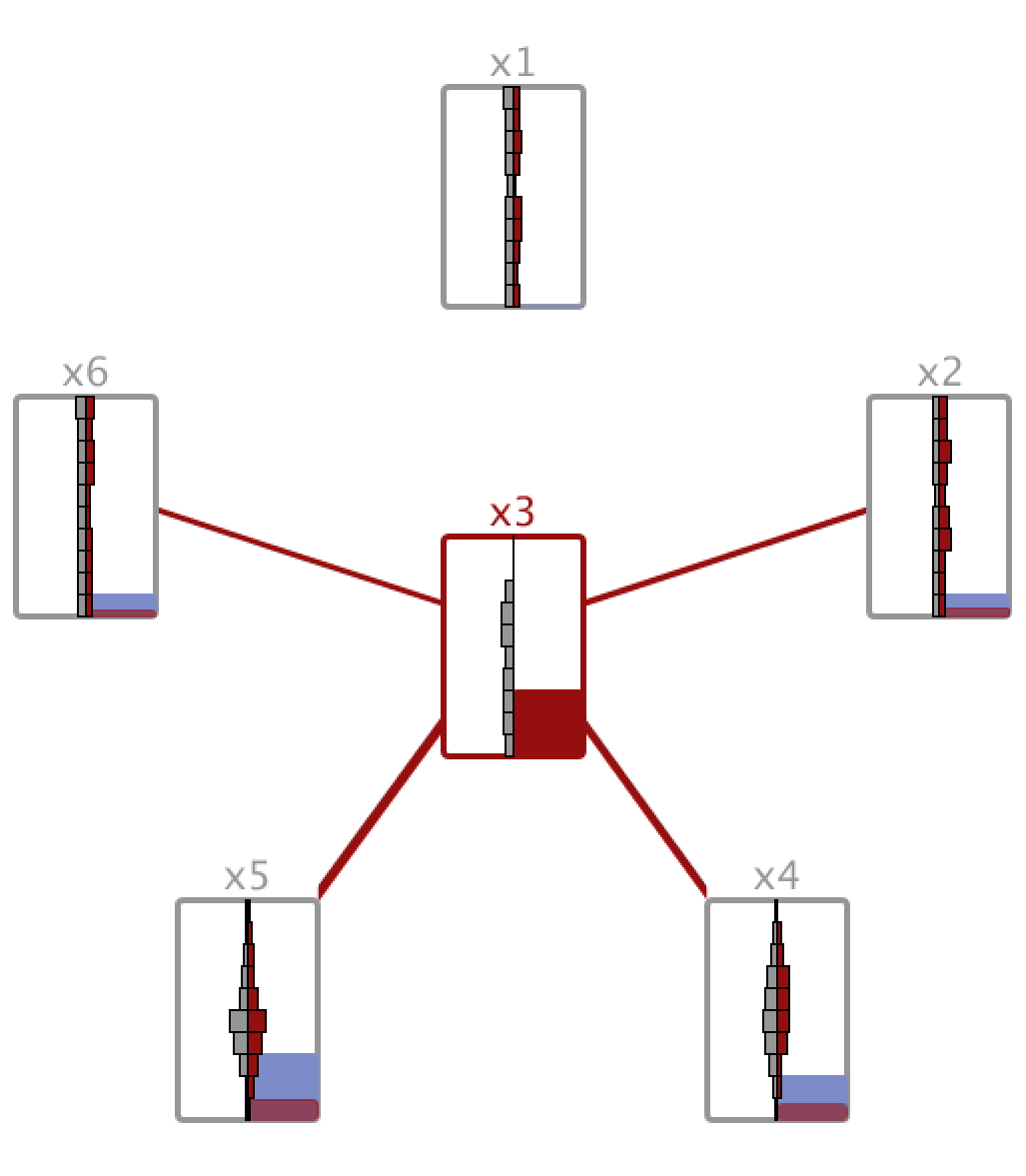}
       \caption{MissiG with radial layout for the same data as in Fig. \ref{Fig:MissiG_Linear_6Var}, with $x3$ selected. Highlighted blocks and bands emphasize JM and red histograms represent CM with missing values in $x3$.}%
       \label{Fig:MissiG_Radial_6Var_x3sel}%
              \vspace{-2mm}
\end{figure}

{\bf Radial layout} is a representation where the glyph of an interactively selected variable of particular interest is highlighted and positioned in the centre of a circle (Fig. \ref{Fig:MissiG_Radial_6Var_x3sel}). The other variables in the data set are positioned on the circumference of the circle, making it possible to analyse the relationships between a single variable of particular interest with respect to all other variables, inspired by the work of Johansson et al.\ \cite{johansson2005}. The width of red bands between glyphs represents the JM with the selected variable, similar to the arcs in the linear layout. The pairwise relationships with the variable of interest may be more easily compared in the radial layout than in the linear, since the distance between the centred variable and other variables is equal and thus band length will not impact perception. The high JM of $x3$ and $x5$ is clearly identifiable in Fig. \ref{Fig:MissiG_Radial_6Var_x3sel}, and it is visible from the width of the red band that the JM of $x3$ and $x4$ is also relatively high. The red histograms are generally mirroring the shapes of corresponding grey histogram, which indicates that there are no strong CM patterns for items with missing values in $x3$.

\begin{figure} [t]%
       \centering
       \subfloat[][PC were missing values are represented below the axes and items with missing values in $x5$ is highlighted in red.]{
       	\includegraphics[width=8.5cm]{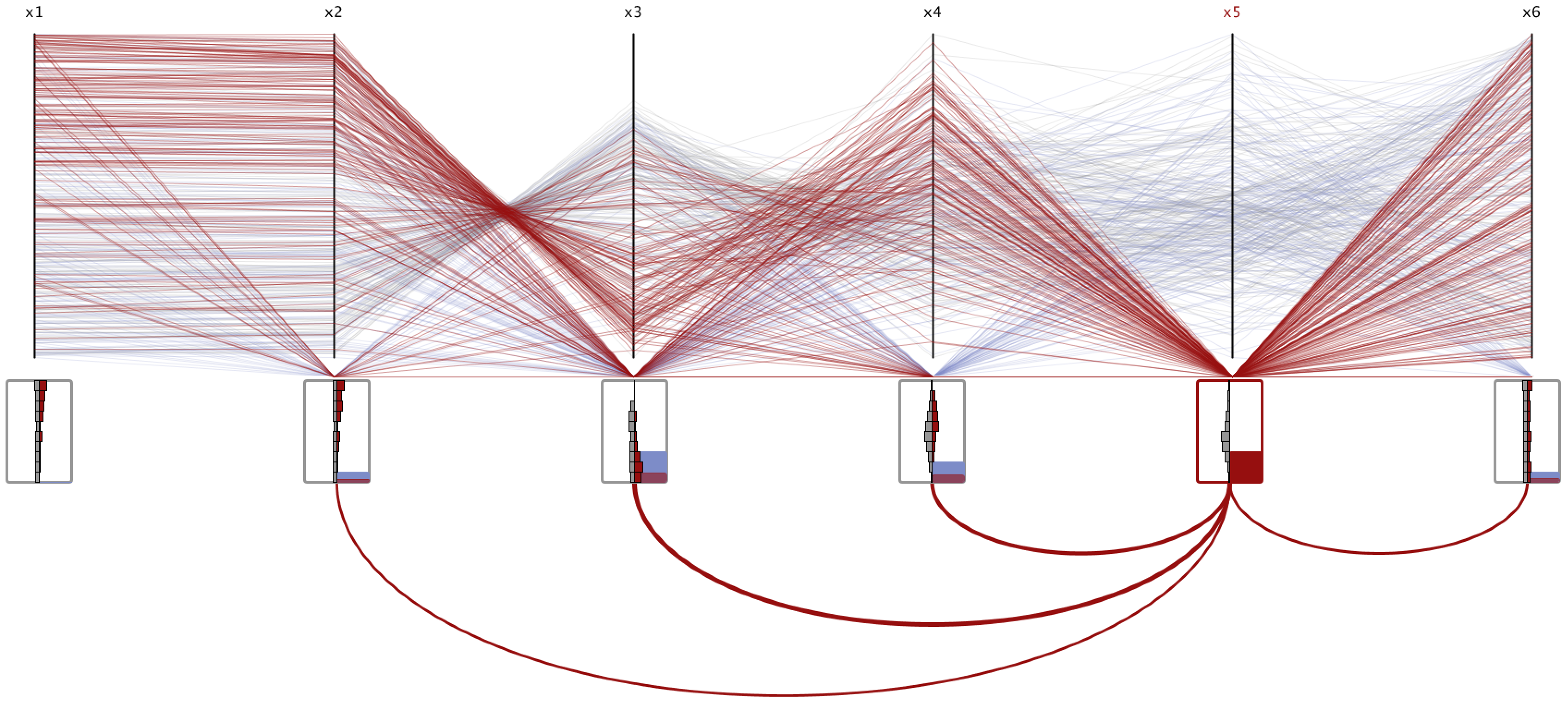}
       	\label{Fig:PC+MissiG_6Var}
       }%
       \qquad
       \subfloat[][Heatmap with missing values represented in red and recorded values represented using grey scale, with dark grey corresponding to low values and light grey corresponding to high values.]{
       	\includegraphics[width=8.5cm]{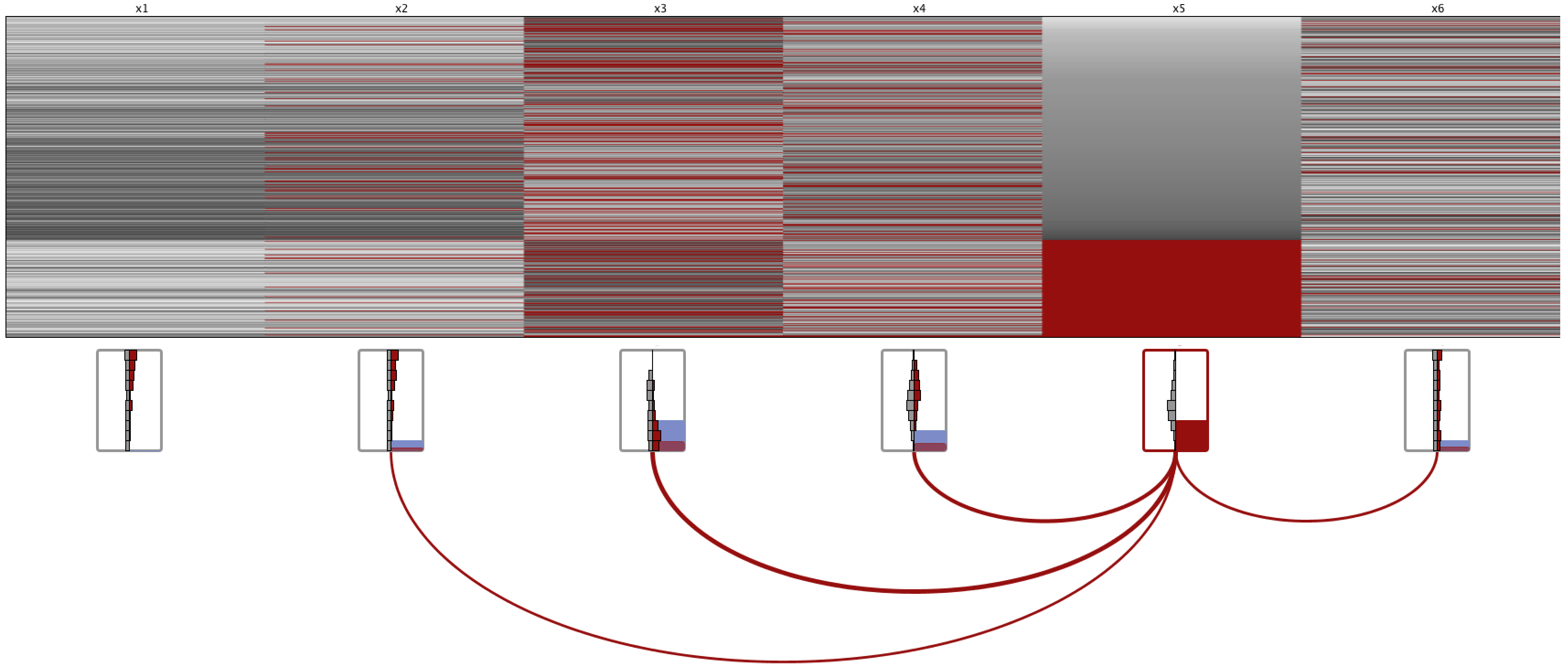}
       	\label{Fig:HM+MissiG_6Var}
       }
       \caption{PC and Heatmap enhanced with MissiG glyphs. Variable $x5$ is selected in both figures.}%
       \label{Fig:PC_HM+MissiG}%
              \vspace{-2mm}
\end{figure}

{\bf Enhancements to existing techniques}. As previously mentioned, the flexibility of the glyph design also makes it usable as an enhancement to existing visualization. Fig. \ref{Fig:PC_HM+MissiG} displays examples where MissiG is used as an enhancement of PC (Fig. \ref{Fig:PC+MissiG_6Var}) and Heatmap (Fig. \ref{Fig:HM+MissiG_6Var}). The PC and Heatmap, as implemented here, already include some representation of missing values. For the PC missing values are represented below the axis, with red highlighting of items with missing values for a selected variable. In the Heatmap, cells with missing values are represented by red colour, while recorded values are represented by grey scale. In the implementation described here, which is mainly focussed on exploration of missing values, MissiG is interactively linked to the other technique through variable selection. Selection of a variable of interest will highlight MissiG as described above. When linked to PC, data items with missing values in the selected variable will be coloured red in PC (see Fig. \ref{Fig:PC+MissiG_6Var}), and when linked to Heatmap the rows will be ordered based on their value in the selected variable (see Fig. \ref{Fig:HM+MissiG_6Var}). The addition of MissiG to these visualization aims to overcome some of the limitations identified in \cite{Fernstad2019}. In Fig. \ref{Fig:PC+MissiG_6Var} it is for instance visible that the JM of $x5$ and $x3$, as well as $x5$ and $x4$ is relatively high, something that likely would have been hard to spot in the PC alone. CM patterns are easier to identify in PC, as confirmed in \cite{Fernstad2019}, but the added MissiG provide a confirmation of patterns such as the relationship between missing values in $x5$ and high values in $x1$, $x2$ and $x4$; particularly in situations where the data is dense with a large amount of missing values. The Heatmap in Fig. \ref{Fig:HM+MissiG_6Var} displays the same data set and selection as the PC. While the Heatmap generally performed well in \cite{Fernstad2019}, particularly for identification of AM and JM patterns, it is still likely that its performance will decline with denser data and high amounts om missing values. For example, the CM relationship between missing in $x5$ and high values in $x4$, and the relationship between missing in $x5$ and high values in $x1$ are equally visible in the MissiG representation, while the former is less perceivable than the latter in the Heatmap, due to $x4$ having a relatively large number of missing values while $x1$ have none. Alternative approaches to enhance visualization methods with MissiG can also be considered. The trade-off between increased understanding of missingness patterns and the potentially increased cognitive burden and possible interference with overall analysis has to be taken into account within the analytical context. In the suggested enhancements the glyph is not overlaying the other visualization, which reduces its visual interference while it requires additional screen-space compared to, for example, overlaying the glyphs on the PC axes. Furthermore, in an analytical context where missing data is not the main focus, it may be appropriate to use a less salient colour than red to highlight missing values and to provide the glyphs as an on-demand feature.

\subsection{Scalability}\label{section::scalability}
\begin{figure} [t]%
       \centering
       \subfloat[][MissiG with linear layout.]{
       	\includegraphics[width=\linewidth]{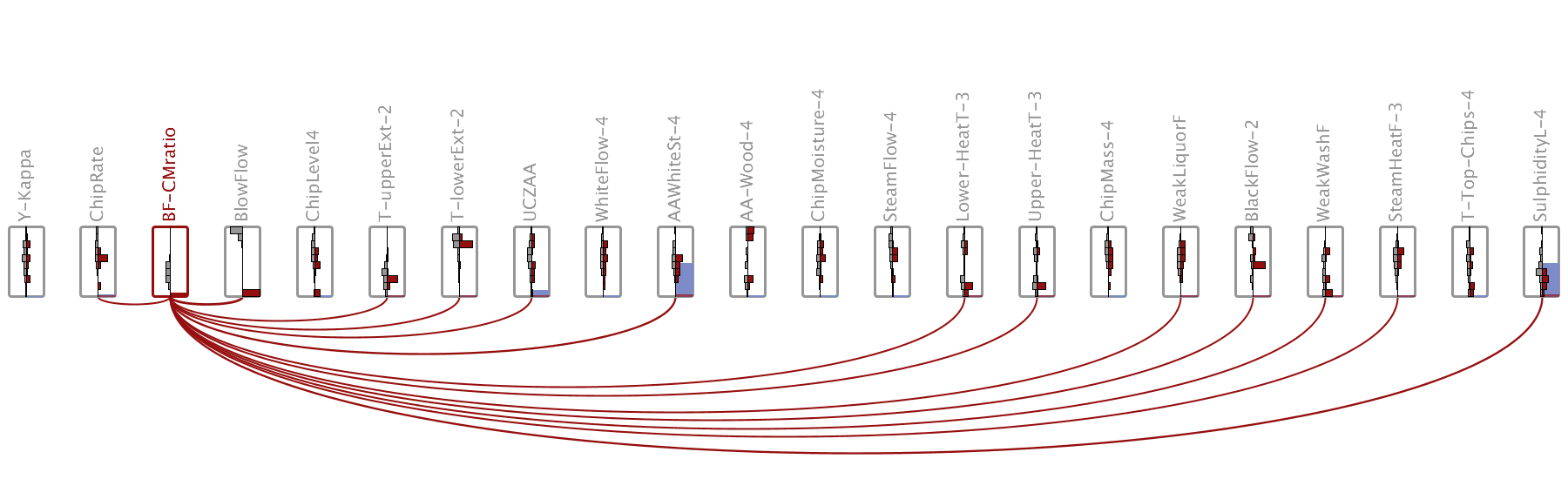}
       	\label{Fig:MissiG_Linear_22Var_B3sel}
       }%
       \qquad
       \subfloat[][PC]{
       	\includegraphics[width=\linewidth]{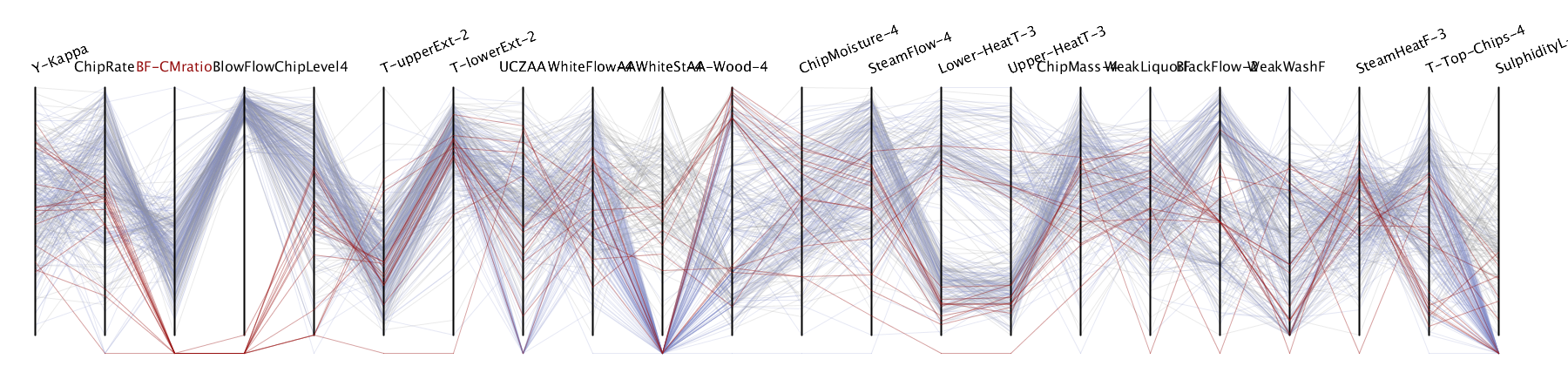}
       	\label{Fig:PC_22Var_B3sel}
       }%
       \qquad
       \subfloat[][Heatmap]{
       	\includegraphics[width=\linewidth]{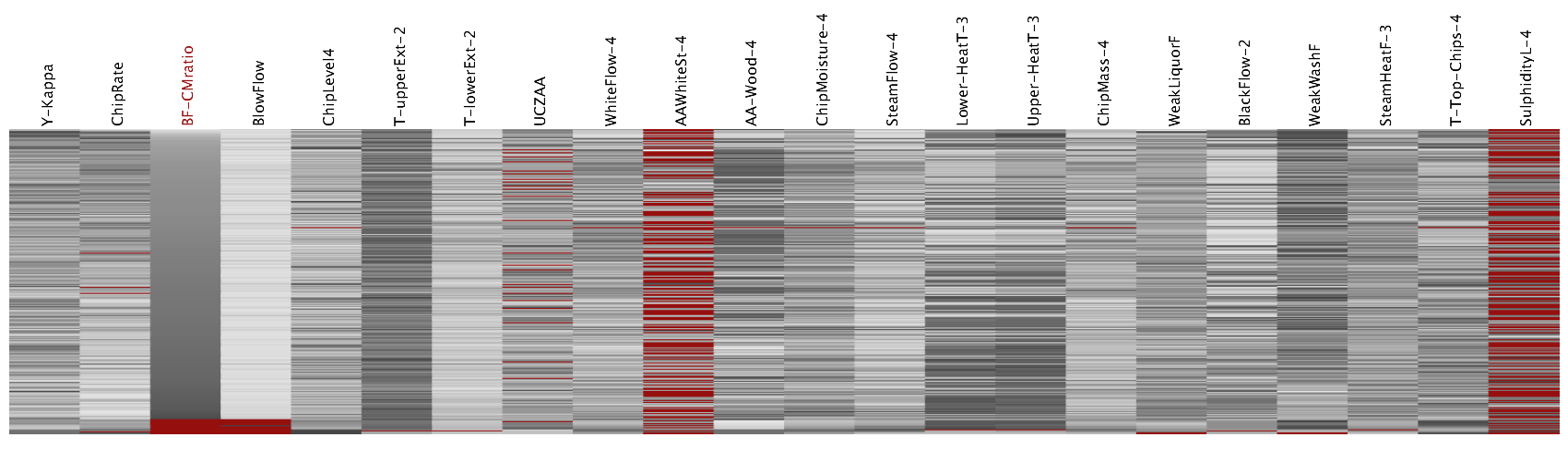}
       	\label{Fig:HM_22Var_B3sel}
       }
       \caption{Visualization of the {\it Kamyr Digester} data set with 22 variables and 301 items. {\it BF-CMratio} (third variable from the left) is selected and highlighted.}%
       \label{Fig:Vis_22Var_B3sel}%
              \vspace{-2mm}
\end{figure}

\begin{figure*} [t]%
       \centering
       \subfloat[][{\it Kamyr Digester} data set with 22 variables and 301 items. {\it BF-CMratio} is selected and positioned in the centre.]{
       	\includegraphics[width=.28\linewidth]{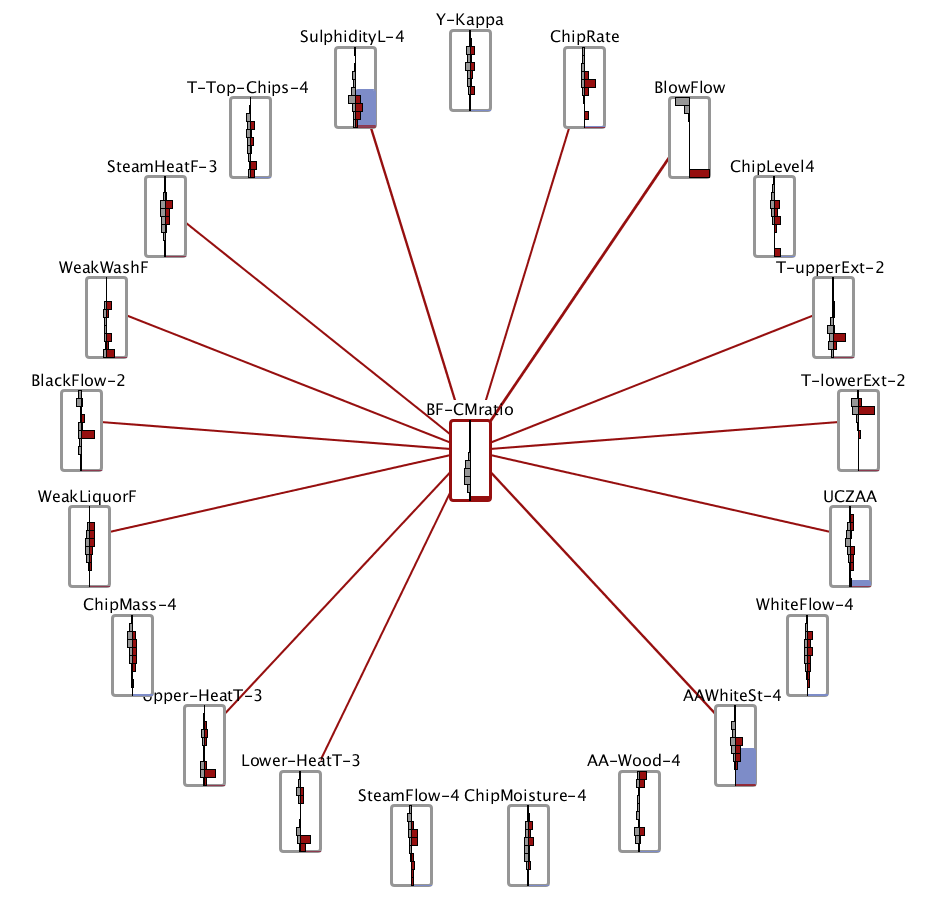}
       	\label{Fig:MissiG_Radial_22Var_B3sel}
       }%
       \qquad
       \subfloat[][{\it Kamyr Digester} data set with 22 variables and 301 items. {\it AAWhiteSt-4} is selected and positioned in the centre.]{
       	\includegraphics[width=.28\linewidth]{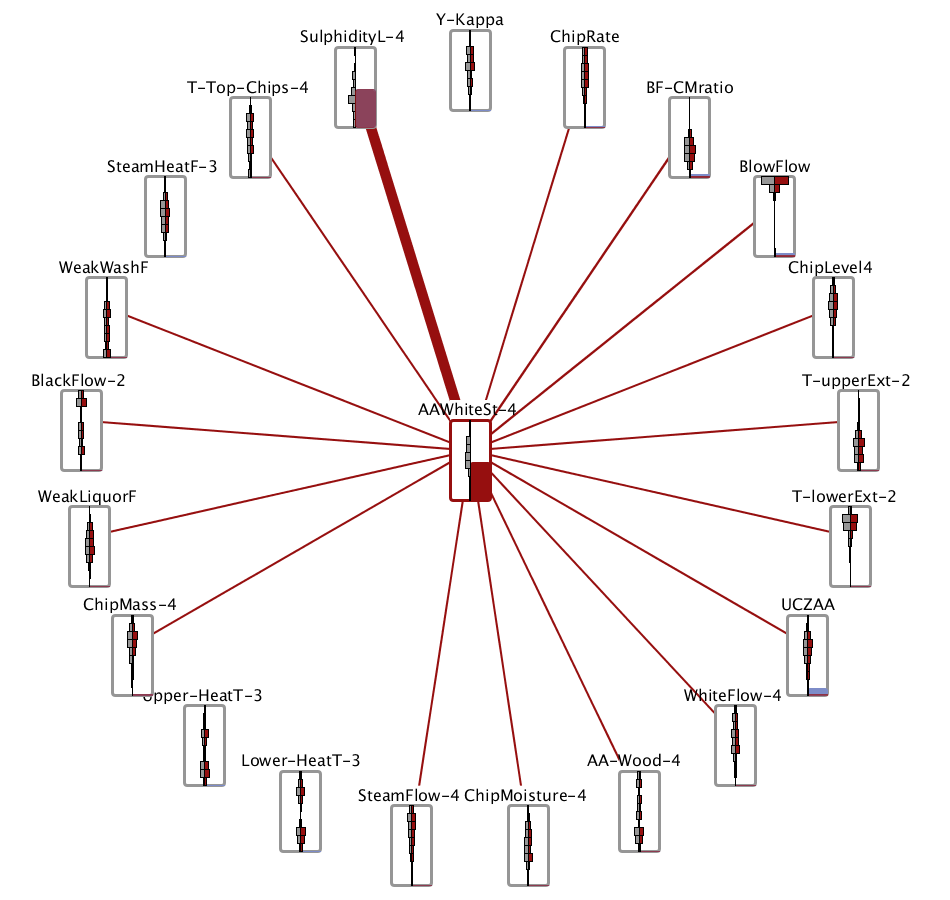}
       	\label{Fig:MissiG_Radial_22Var_B10sel}
       }%
       \qquad
       \subfloat[][The {\it Communities and Crime} data set with the {\it PolicBudgPerPop} variable selected and positioned in the centre.]{
       	\includegraphics[width=.28\linewidth]{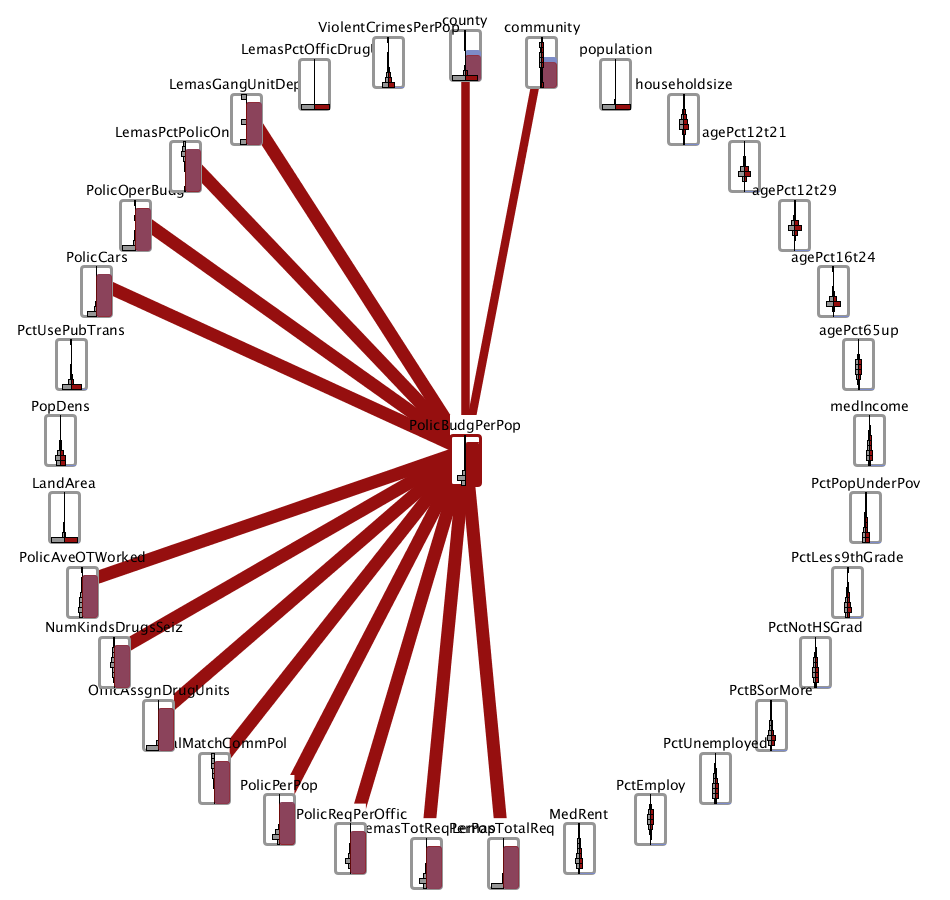}
       	\label{Fig:MissiG_Radial_34Var}
       }
       \caption{Visualization of two data sets using MissiG with radial layout.}%
       \label{Fig:MissiG_Radial_22_34}%
              \vspace{-2mm}
\end{figure*}

\begin{figure} [t]%
       \centering
       \subfloat[][MissiG with linear layout.]{
       	\includegraphics[width=\linewidth]{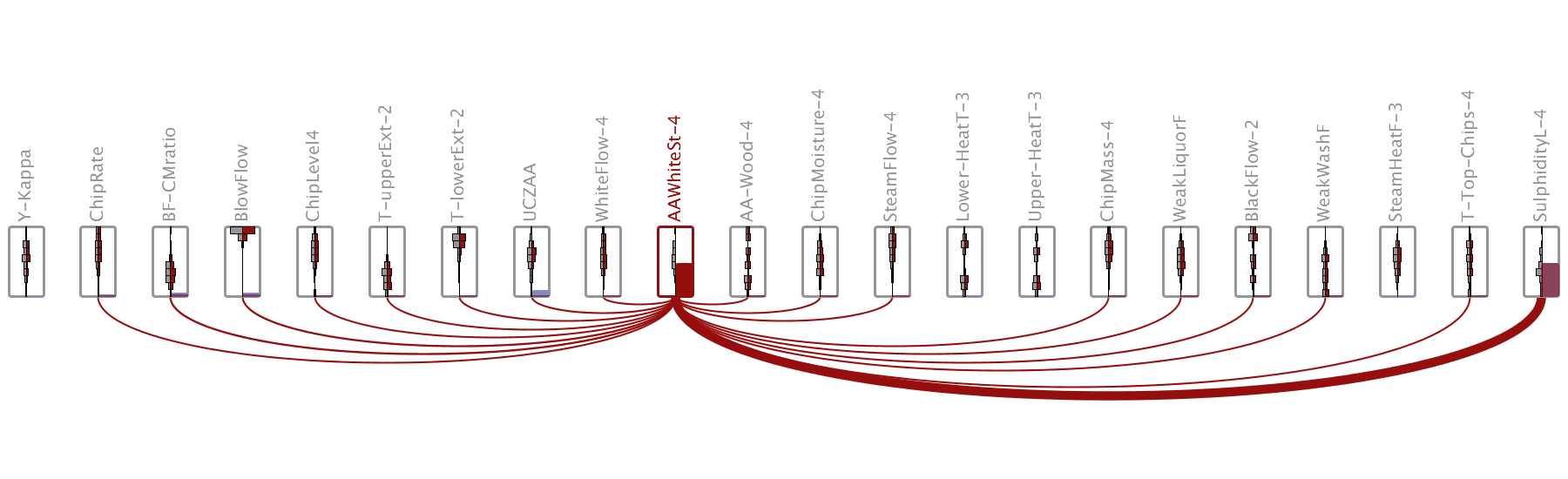}
       	\label{Fig:MissiG_Linear_22Var_B10sel}
       }%
       \qquad
       \subfloat[][PC]{
       	\includegraphics[width=\linewidth]{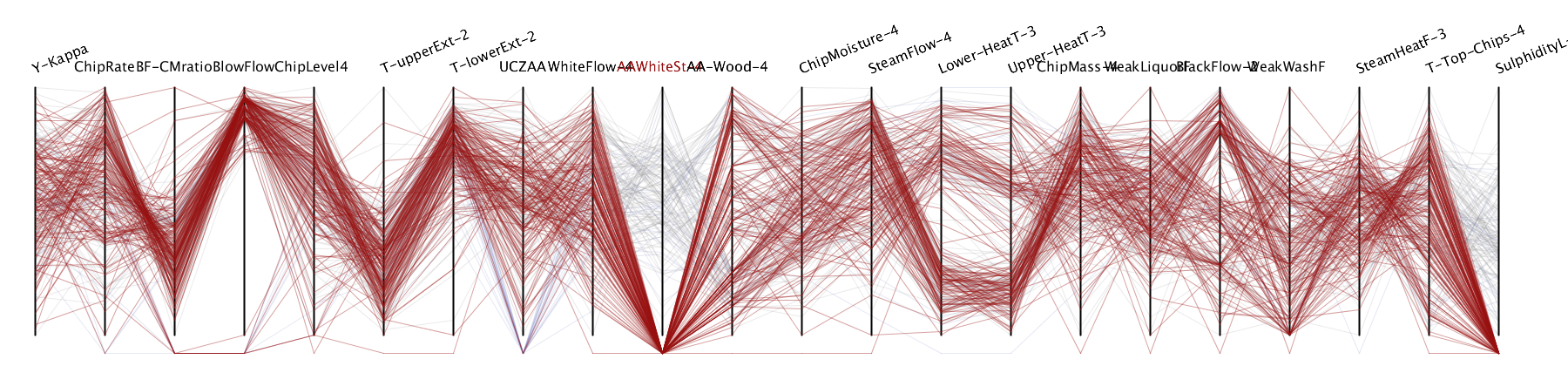}
       	\label{Fig:PC_22Var_B10sel}
       }%
       \qquad
       \subfloat[][Heatmap]{
       	\includegraphics[width=\linewidth]{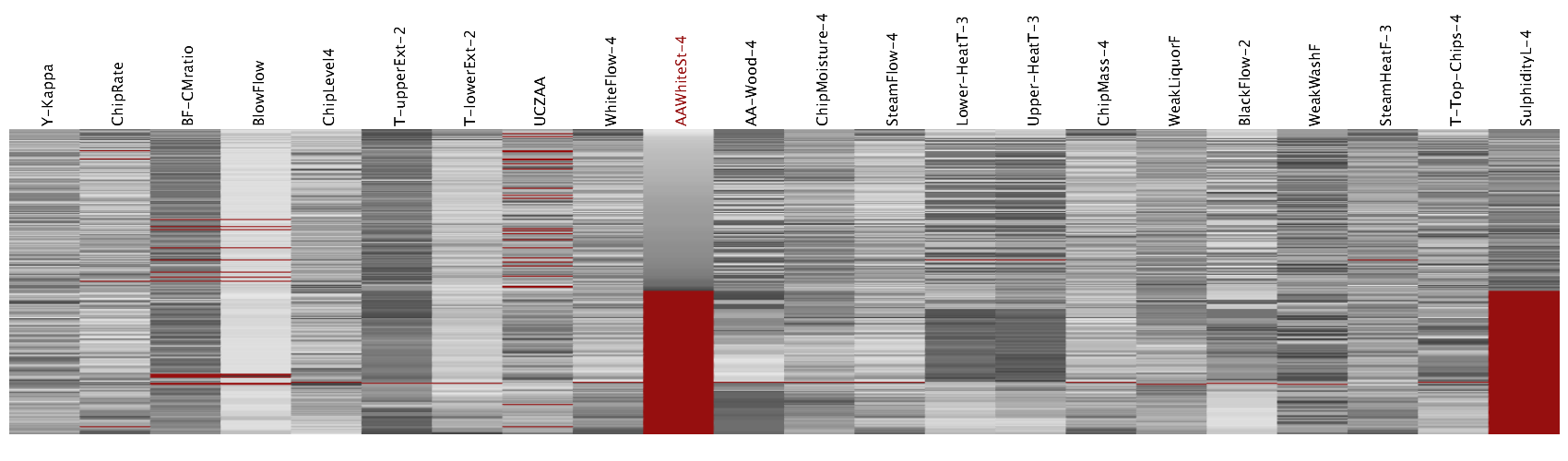}
       	\label{Fig:HM_22Var_B10sel}
       }
       \caption{Visualization of the {\it Kamyr Digester} data set with 22 variables and 301 items. {\it AAWhiteSt-4} (tenth variable from the left) is selected and highlighted.}%
       \label{Fig:Vis_22Var_B10sel}%
              \vspace{-2mm}
\end{figure}

Due to the relative simplicity of the glyph design and the representation of summary patterns rather than features of individual data items, the scalability of MissiG is comparable to or even better than the scalability of common multivariate visualization methods such as PC, Heatmap or Scatterplot matrix. The visual complexity of a single glyph will not grow with an increasing number of data items and, hence, from a perception and interpretation point of view it will not matter if the data set includes a few hundred items or a million items. On the other hand, with a multivariate MissiG layout, an increasing number of variables will require additional visual objects to be drawn and impact the usability of the visualization, similarly to how it impacts common multivariate visualization methods. The exact number of variables that can be visualized effectively depends not only on the available display space, but also on the missingness patterns in the data as well as which layout and visualization options are being used. A data set where only a small number of variables have missing values will likely result in a less cluttered display than a data set with missing values in most variables; and a linear display where all JM arcs are being displayed (as in Fig. \ref{Fig:MissiG_Linear_6Var_allArcs}) will often result in a more cluttered display than if only JM arcs of a selected variable is displayed (as in Fig. \ref{Fig:MissiG_Linear_6Var_x5sel}). Fig. \ref{Fig:Vis_22Var_B3sel}, \ref{Fig:MissiG_Radial_22_34}, \ref{Fig:Vis_22Var_B10sel} and \ref{Fig:Vis_34Var} show the use of MissiG, PC and Heatmap for differently sized data sets with different missingness distributions and patterns. As with other mutlivariate visualization techniques, approaches to deal with more complex and higher dimensional datasets may for example include simplified glyph design with details on demand, dimensionality reduction and quality metric approaches to aid identification of the most interesting missingness patterns.

\section{Examples of Pattern Identification}\label{section::usecase}
This section provides examples of how MissiG can be useful for identifying missingness patterns. Larger versions of the figures (\ref{Fig:Vis_22Var_B3sel} - \ref{Fig:Vis_34Var}) are provided in the supplemental material. The linear and radial MissiG layouts will be compared with PC and Heatmap. In  PC, missing values are represented at a point below the axis, and items with missing values for a selected variable are highlighted in red. The Heatmap use red colour to represent missing values and grey scale to represent recorded values with light grey corresponding to high values. The rows in the Heatmap are ordered based on their values for a selected variable. Two public data sets with missing values are used for the examples, the {\it Kamyr Digester} data set \cite{Dayal1994} which contains 22 variables (observation date was removed from the data), 301 items and has 352 (5.6\%) missing values, and the {\it Communities and Crime} data set \cite{UCI}, which in full contains 128 variables and 1994 data items. For the purpose of the examples here, 34 variables were randomly selected from the {\it Communities and Crime} data, with a total of 24126 (55.2\%) missing values.

Fig. \ref{Fig:Vis_22Var_B3sel} and \ref{Fig:MissiG_Radial_22Var_B3sel} displays the {\it Kamyr Digester} data set with the {\it BF-CMratio} variable selected. It is visible in the MissiG plots (\ref{Fig:MissiG_Linear_22Var_B3sel} and \ref{Fig:MissiG_Radial_22Var_B3sel}) and the Heatmap (\ref{Fig:HM_22Var_B3sel}) that there is a strong JM between {\it BF-CMratio} and {\it BlowFlow} (third and fourth from left). This can to some extent also be identified in the PC (\ref{Fig:PC_22Var_B3sel}) through the red line below the third and fourth axis from the left, but it is considerably harder to appreciate the amount of jointly missing values in PC. Looking at the histograms in the MissiG glyph of {\it SulphidityL-4} (rightmost glyph in \ref{Fig:MissiG_Linear_22Var_B3sel} and upper left in \ref{Fig:MissiG_Radial_22Var_B3sel}), it is visible that the shapes of the grey and red histograms differ, with the red histogram being denser towards lower values. This indicates a potential CM between missing values in the selected {\it BF-CMratio} and low recorded values in {\it SulphidityL-4}. The same pattern is considerably harder to identify in the PC (\ref{Fig:PC_22Var_B3sel}) and in the Heatmap (\ref{Fig:HM_22Var_B3sel}) due to the high number of missing values in {\it SulphidityL-4} which largely masks the patterns of recorded values. These JM and CM patterns are clearly not random patterns, and if not already known it may prompt the analyst to further examine the connection between these variables and investigate potential issues in the data collection or pre-processing steps.

Fig. \ref{Fig:Vis_22Var_B10sel} and \ref{Fig:MissiG_Radial_22Var_B10sel} display the same data set as in Fig. \ref{Fig:Vis_22Var_B3sel} but with {\it AAWhiteSt-4} (tenth variable from the left) selected in all views. Starting with CM patterns, it is visible in Fig. \ref{Fig:MissiG_Linear_22Var_B10sel} and \ref{Fig:MissiG_Radial_22Var_B10sel} that the shapes of the red histograms are very similar to the shapes of corresponding grey histograms. This indicates that items with missing values in {\it AAWhiteSt-4} are randomly distributed across the recorded values of other variables, which may aid the choice of imputation method. These patterns are harder to visually verify in the PC and Heatmap (Fig. \ref{Fig:PC_22Var_B10sel} and \ref{Fig:HM_22Var_B10sel}), and in particular when patterns occur such as the dark grey blocks in variables {\it Lower-HeatT-3} and {\it Upper-HeatT-3} in the Heatmap (column four and five to the right of the selected {\it AAWhiteSt-4}), which are a result of the two density peaks in the variables (as visible from the histograms in \ref{Fig:MissiG_Linear_22Var_B10sel}) combined with the ordering of items which in this figure is based on values in {\it AAWhiteSt-4}. While investigating JM, two potentially interesting relationships are identified in Fig. \ref{Fig:MissiG_Linear_22Var_B10sel} and \ref{Fig:MissiG_Radial_22Var_B10sel}. Firstly, from the red block in the rightmost variable ({\it SulphidityL-4}), which completely overlaps the blue block, it appears that all items that are missing in {\it AAWhiteSt-4} are also missing in {\it SulphidityL-4}. Both {\it AAWhiteSt-4} and {\it SulphidityL-4} have around 50\% of values missing, which means that around 25\% of items should be jointly missing if the values were missing completely at random, hence, the high JM indicates a non-random missingness pattern suggesting that the missingness in these variables and its cause should be investigated in conjunction. The pattern is easily identifiable also in the Heatmap (Fig. \ref{Fig:HM_22Var_B10sel}) through the blocks of red rows, and in the PC (Fig. \ref{Fig:PC_22Var_B10sel} where the items with missing values in {\it AAWhiteSt-4} (red lines) all intersects below the {\it AAWhiteSt-4} axis, although the percentage missing is hard to read from the PC. The second JM pattern that can be identified in the MissiG representations is that the JM between {\it AAWhiteSt-4} and {\it UCZAA} (two steps left of {\it AAWhiteSt-4} in \ref{Fig:MissiG_Linear_22Var_B10sel}) is lower than expected from a random pattern. With half of values missing for {\it AAWhiteSt-4} it is expected that the red block in {\it UCZAA} would be half the size of the blue block in {\it UCZAA} if the missingness was random, but the red block is considerably smaller and hence indicates a non-random pattern. The same pattern can be spotted in the Heatmap, although less obvious, while it is considerably harder in PC since it does not clearly represent the frequency of missing values.

\begin{figure} [t]%
       \centering
       \subfloat[][Linear MissiG with the {\it Community} variable (second from left) selected. ]{
       	\includegraphics[width=\linewidth]{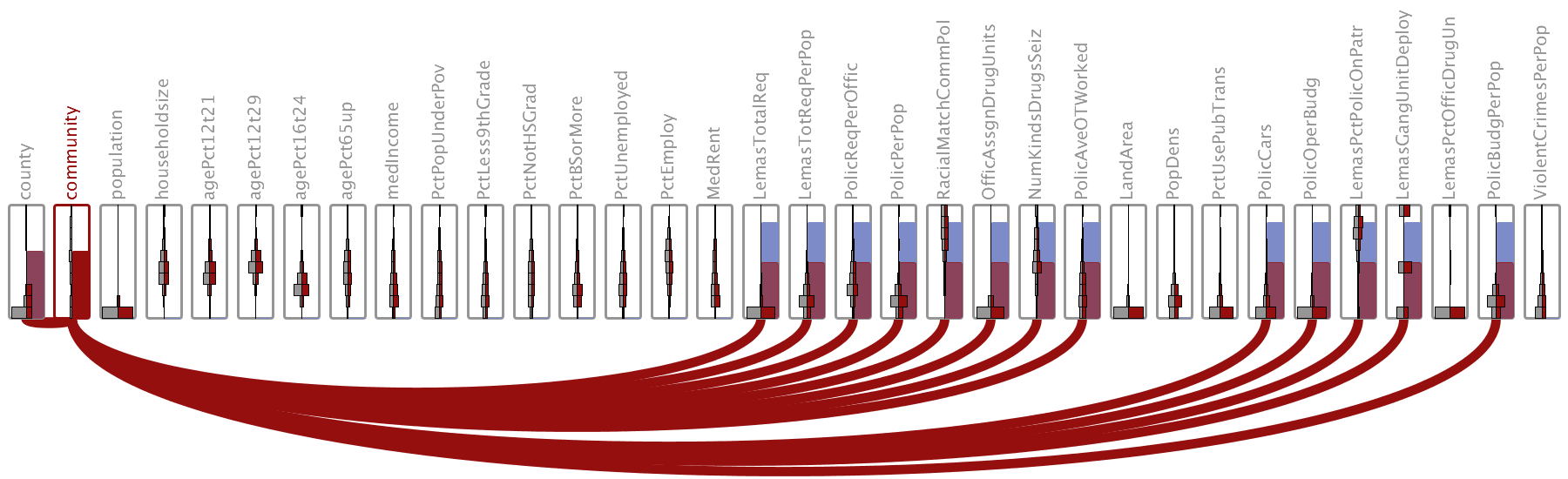}
       	\label{Fig:MissiG_Linear_34Var}
       }%
       \qquad
       \subfloat[][PC where the {\it Community} variable (second from left) is selected.]{
       	\includegraphics[width=\linewidth]{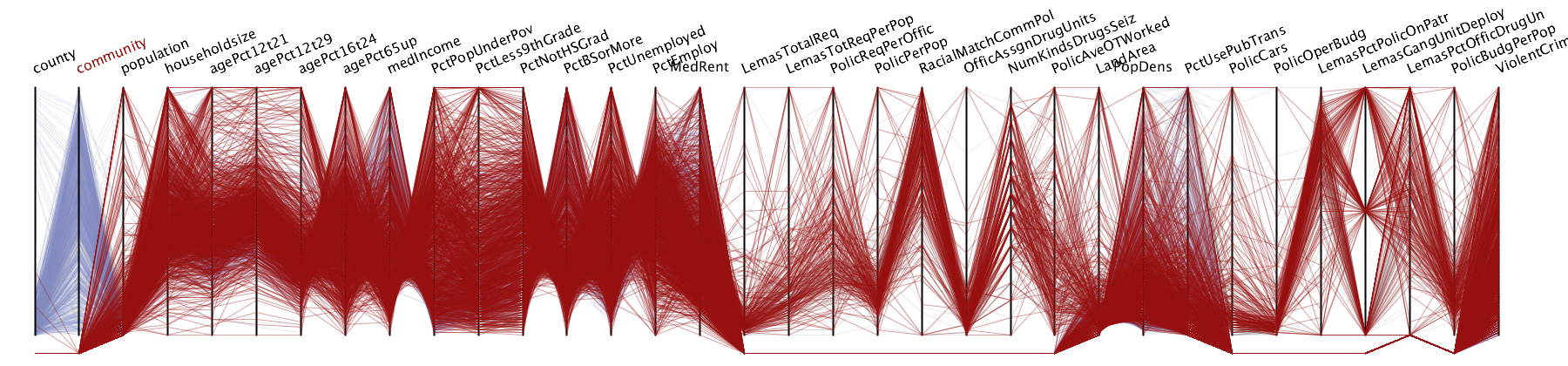}
       	\label{Fig:PC_34Var}
       }%
       \qquad
       \subfloat[][Heatmap with {\it PolicBudgPerPop} variable (second from right) selected.]{
       	\includegraphics[width=\linewidth]{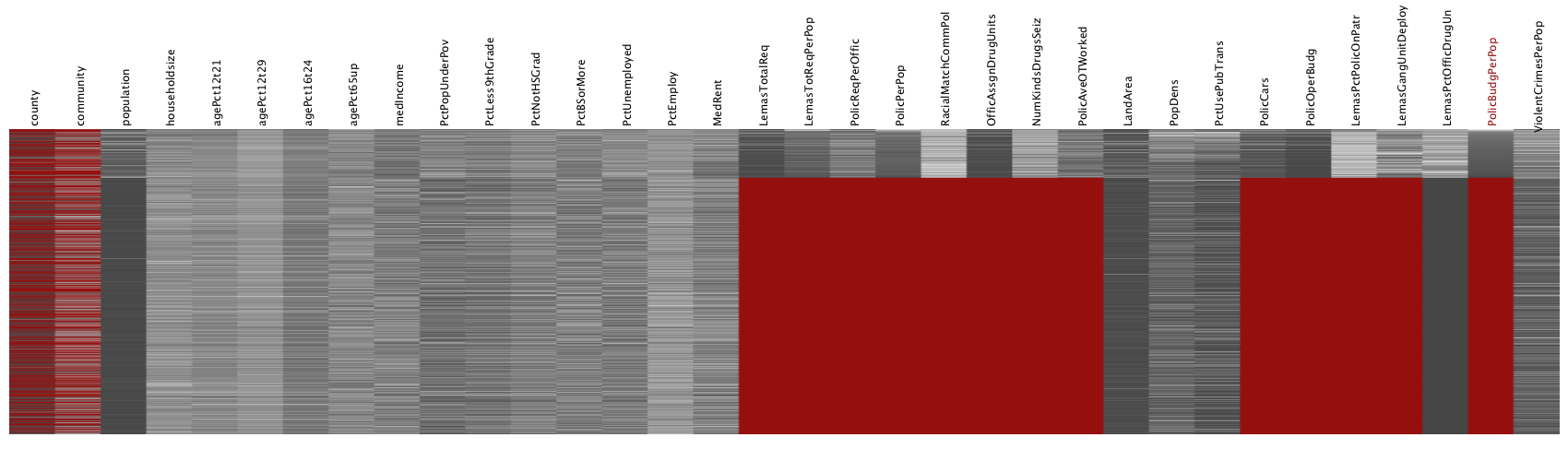}
       	\label{Fig:HM_34Var}
       }
       \caption{Visualization of 34 variables and 1994 data items from the {\it Communities and Crime} data set. }%
       \label{Fig:Vis_34Var}%
              \vspace{-2mm}
\end{figure}

Fig. \ref{Fig:Vis_34Var} and \ref{Fig:MissiG_Radial_34Var} displays 34 variables from the {\it Communities and Crime} data set. The {\it Community} variable (second from left) is selected in Fig. \ref{Fig:MissiG_Linear_34Var} and \ref{Fig:PC_34Var}, and some patterns related to distribution of missing values are easily identified in the linear MissiG layout. Only 15 variables have missing values. Of these, the two leftmost have around 50\% of values missing, as visible from the blocks in the lower right part of corresponding glyphs, while the remaining 13 variables have nearly 85\% missing, which can be seen from the blue blocks that are nearly as high as the full height of the glyphs. In the PC (Fig. \ref{Fig:PC_34Var}) it can be seen from the high number of red lines that the selected variable has a relatively high AM, but the AM in other variables is hard to appreciate. It can be seen from the two leftmost glyphs in Fig. \ref{Fig:MissiG_Linear_34Var} that nearly all items with missing values for {\it Community} also have missing values for {\it County}, which is also visible in the PC (\ref{Fig:PC_34Var}) from the small number of red lines linking from missing values in {\it Community} (second axis) to recorded values along the leftmost {\it County} axis. It is also clearly visible in the MissiG representation that approximately half of the items that have missing values in the 13 variables to the right, also have missing values for the {\it Community} variable, since the red blocks in the 13 variables is around half the height of the blue blocks, which can be expected given the AM of the individual variables. While it is visible in the PC that there is JM between {\it Community} and the 13 variables to the right, it is difficult to estimate the size of JM.

In Fig. \ref{Fig:MissiG_Radial_34Var} and \ref{Fig:HM_34Var} the {\it PolicBudgPerPop} variable is selected (second from right in Heatmap and centre in radial MissiG). It is clear from both figures that a majority of values (around 85\%) are missing from this variable, and that more or less all of the items with missing values in {\it PolicBudgPerPop} also have missing values in 12 other variables, as visible from the red blocks in the MissiG completely covering the blue blocks, and from the red blocks in the Heatmap where rows are ordered by values {\it PolicBudgPerPop}, this indicates a relationship between missing in {\it PolicBudgPerPop} and missing in the other 12 variables that is not occurring at random. If this JM pattern had been random, a JM rate of around 85\% would be expected rather than 100\%. In Fig. \ref{Fig:MissiG_Radial_34Var} it is however visible that the JM for the selected {\it PolicBudgPerPop} and the {\it County} and {\it Community} variables (top and top right glyph) is likely around 85\%, since a part of the blue block is visible above the red block in the {\it County} and {\it Community} blocks, and that the grey and red histograms for other variables have similar shapes, which indicate a random relationship with missing values in {\it PolicBudgPerPop}. These patterns are not as easily detected in the Heatmap, where it for instance is hard to appreciate the amount of jointly missing values with {\it County} and {\it Community}. While the high AM suggest that any imputation would introduce too much bias, the knowledge of JM patterns can help understand data collection issues that may need to be addressed.

\section{Evaluation of the Missingness Glyph}\label{section:evaluation}
Two usability studies were carried out to establish the usability of MissiG. The first was designed using interactive visualization in the lab, while the second was an online study comparing static visual representations. The following section will describe the studies and their results in detail. 

\subsection{Visualization Methods}
Six visualization methods were compared in the studies, including different MissiG layouts and extensions. There was a slight variation in the MissiG design between the first and second study, as the glyph was improved after the first study. In the initial design (Fig. \ref{Fig:MissiG_old}) the amount missing blocks were stretched across the full width of the glyph, and histograms were squeezed into the space available above the blocks, resulting in different histogram heights for variables and difficulties comparing histograms across variables. As shown in Fig. \ref{Fig:MissVisDescription} the amount missing blocks in the final design only cover half the width, to allow more space for the histograms, and use opacity and black borders to increase visibility of the part of the red CM histogram that overlap the blocks. The visualization methods compared were:

\begin{figure} [t]%
\centering
       	\includegraphics[width=5.5cm]{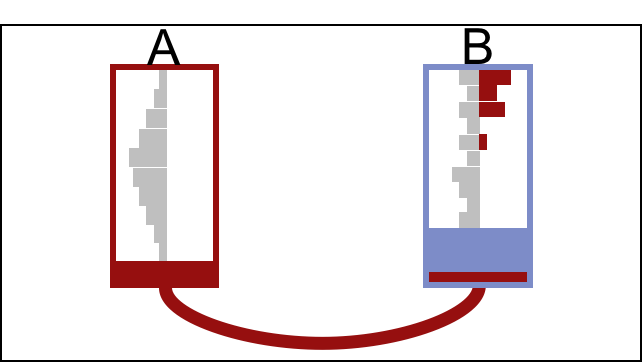}
       \caption{The initial MissiG design, with amount missing blocks across the width of the glyph and histograms above the blocks. }%
       \label{Fig:MissiG_old}%
       \vspace{-2mm}
\end{figure}

\noindent
\textbf{Linear MissiG (MissiG-L)}: The linear layout of MissiG with display of JM arcs only for the selected variable, as in Fig. \ref{Fig:MissiG_Linear_6Var_x5sel}.

\noindent
\textbf{Radial MissiG (MissiG-R)}: The radial layout of MissiG, where the selected variable is positioned in the centre, as in Fig. \ref{Fig:MissiG_Radial_6Var_x3sel}. This layout was only included in the second study.

\noindent
\textbf{Heatmap (HM)}: Heatmap where missing values are represented by red cells, and recorded values are represented in grey scale with dark corresponding to low values and light corresponding to high values.

\noindent
\textbf{Heatmap with MissiG (HM+MissiG)}: Heatmap with the same colouring as above but enhanced with MissiG, as in Fig. \ref{Fig:HM+MissiG_6Var}.

\noindent
\textbf{Parallel Coordinates (PC)}: PC where missing values are represented by polylines intersecting a point below the axis, and items with missing values for a selected variable is represented by red polylines.

\noindent
\textbf{Parallel Coordinates with MissiG (PC+MissiG)}: PC with missing value representation as above but enhanced with MissiG, as in Fig. \ref{Fig:PC+MissiG_6Var}.

\subsection{Hypotheses and Tasks} \label{section:tasks}
Based on the results in \cite{Fernstad2019}, the evaluations were designed to test the following hypotheses:
\begin{itemize}
    \item[\textbf{H1}] The MissiG glyphs will perform better than PC and better or equally well as Heatmap for AM tasks.
    \item[\textbf{H2}] PC+MissiG will perform better than PC for AM tasks.
    \item[\textbf{H3}] HM+MissiG will perform better or equally well as Heatmap for AM tasks.
    \item[\textbf{H4}] The MissiG glyphs will perform better than PC and better or equally well as Heatmap for JM tasks.
    \item[\textbf{H5}] PC+MissiG will perform better than PC for JM tasks.
    \item[\textbf{H6}] HM+MissiG will perform better or equally well as Heatmap for JM tasks.
    \item[\textbf{H7}] The MissiG glyphs will perform better than Heatmap and better or equally well as PC for CM tasks.
    \item[\textbf{H8}] PC+MissiG will perform better or equally well as PC for CM tasks.
    \item[\textbf{H9}] HM+MissiG will perform better than Heatmap for CM tasks.
    \item[\textbf{H10}] Overall, MissiG will be the most preferred visualization method by participants.
\end{itemize}

Tasks were aimed to address different aspects of the three missingness patterns. This included approximation of percentage of missing values in a variable; identification of the variable with most missing values; identification of the variable pair that has highest joint missingness; comparison of difference in joint missingness between variable pairs; evaluation of trends in recorded data that possibly relates to missing values (such as: items with missing values in variable A tend to have high recorded values in variable B); and identification of differences between the general data distribution and the distribution of items that have missing values in some variable. To cover this as broadly as possible, two different questions were defined for each missingness patterns, with multiple choice style answers provided. The questions were defined as follows:

\begin{itemize}
    \item[\textbf{AM1:}] Approximately how much data is missing in variable $X$?
    \item[\textbf{AM2:}] Which of the following variables have the highest number of missing values?
    \item[\textbf{JM1:}] With which of the following variable does variable $X$ have the highest joint missingness? 
    \item[\textbf{JM2:}] Is the joint missingness of variable $X$ and $Y$ higher than the joint missingness of variable $X$ and $Z$?
    \item[\textbf{CM1:}] Which of the following trends is most related to missing values in variable $X$?
    \item[\textbf{CM2:}] Which image displays the highest number of variables with a clear difference between the general data distribution and the distribution of items that are missing in the selected variable
\end{itemize}

For both studies, the performance of the visualization methods was analysed in terms of accuracy and response time when completing the tasks. Based on the study hypothesis, three sets of analysis were relevant for each missingness pattern: 1) MissiG vs HM vs PC; 2) HM vs HM+MissiG; and 3) PC vs PC+MissiG. For the first set of significance testing, where more than two visualization methods were compared, one way ANOVA with repeated measures was used when the result data were normally distributed. If data were not normally distributed the Friedman test was used, followed by post-hoc tests for pairwise comparison to identify for which combinations of visualization methods the performance was significantly different, using Wilcoxon signed rank test with a Bonferroni correction applied resulting in a significance level set at $p < 0.017$ for the first study (comparing three methods), and at $p < 0.0125$ for the second study (comparing four methods). For the second and third set of significance testing, where pairs of visualization methods were compared, a dependent t-Test was used for normally distributed data, while Wilcoxon signed rank test was used for non-normally distributed data.

\subsection{First Study}
A study with 15 participants was conducted to evaluate the performance of an initial implementation of the MissiG visualization. This study compared MissiG-L with Heatmap and PC, as well as comparing standard Heatmap and PC with versions enhanced with MissiG glyphs, using an interactive environment allowing for highlighting of glyphs and polylines in PC, and sorting of rows in Heatmap (as described in Section \ref{Sec:layouts}). Three of the above questions (\textbf{AM1}, \textbf{JM1} and \textbf{CM1}) where used in the study.

\subsubsection{Experimental Design and Procedure}
The experiment was designed as a within-subject study with visualization method as factor. Each participant performed 45 tasks and equally many tasks were performed for each visualization method and pattern using the data sets described in \ref{section:FirstStudy_data}. No data set was used more than once per participant. Performance was measured in terms of accuracy and response time when performing the tasks. Ethical approval was received prior to the study.

\begin{figure} [t]%
\centering
       	\includegraphics[width=8cm]{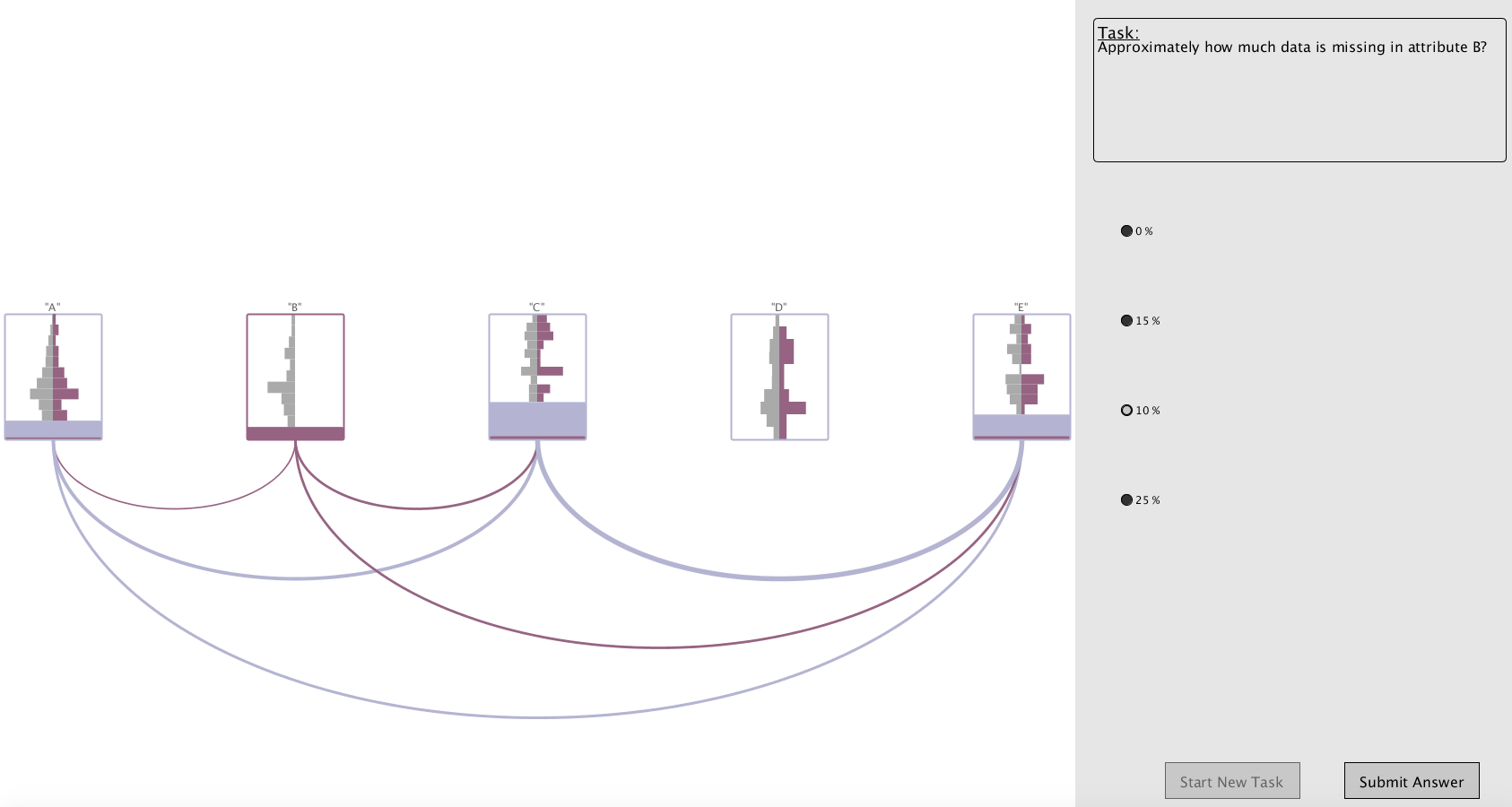}
       \caption{The interface of the first study, with interactive visualization on the left and multiple choice options to the right.}%
       \label{Fig:Study1_interface}%
       \vspace{-2mm}
\end{figure}

The study was conducted individually in a controlled setting using a 15-inch MacBook Pro and an external screen where the study interface was displayed as a fixed size 1700x920px window. An initial scripted presentation was used to ensure that all participants possessed the basic knowledge needed to interpret the visual representations and understand the missingness patterns and tasks. This was followed by a training period including a small number of test tasks using the different visualization methods. The training was used as a means for the participants to become familiar with the tasks, visualization methods and experimental environment. For the experimental phase the tasks and visualization methods were counterbalanced using a Latin-square procedure \cite{Graziano1993}, resulting in a unique ordering for each participant and, hence, reducing the potential learning impact on the results. The test environment (Fig. \ref{Fig:Study1_interface}) consisted of two panels, one displaying one of the interactive visualization methods representing a study data sets, and the other displaying the task and multiple choice answers, along with buttons for submitting the answer and for displaying a new question. Response time was measured from when a question was displayed until the answer was submitted, allowing the participants to take a break before displaying a new question. The answers provided and response times were stored in log files, which were later used to analyse the results. The experimental phase was followed by a short questionnaire collecting information about the participants and their previous experience of data analysis, visualization and missing data, as well as information about which visualization method they found easiest, hardest and preferred to use.

\subsubsection{Data}\label{section:FirstStudy_data}
Three publicly available data sets where used and modified through controlled removal of values, to maintain realistic data structures while controlling the missingness patterns in the data. A total of 45 data sets were generated, 15 based on the {\it User Knowledge Modelling} (UKM) data set \cite{Kahraman2013} with 5 variables and 403 items, 15 based on the {\it Concrete Compressive Strength} data set \cite{Yeh1998} with 9 variables and 1030 items, and 15 based on the {\it Parkinsons} data set \cite{Little2007} with 23 variables and 197 items. Varying levels of uniformly distributed noise, with a noise level between 1\% and 15\%, was randomly added, and the data sets were separated into three groups, one for each missingness pattern. Missingness patterns were created by replacing numerical values with a $NaN$ string, with between 0\% and 40\% of values removed from each variable. The structure of missingness in variables was defined using a similar approach to Fernstad \cite{Fernstad2019}. The variable names of the original data were replaced by letters to avoid impact of preconceptions based on variable names.

\subsubsection{Results}
15 participants finished the study, 2 were female and 13 male. The biggest age group among participants was 25-34 years (46.7\%), followed by 45-54 (26.7\%), 35-44 (20\%) and 18-24 (6.7\%). Participants were asked to rank their level of experience of 1) visualization methods, 2) data analysis, and 3) missing data, using 5 point likert scales ranging from No prior experience (1) to Professional (5). 46.7\% ranked their experience of visualization methods high (4 or 5), while almost equally many (40\%) ranked their visualization experience low (1 or 2). A clear majority (86.7\%) ranked their experience of data analysis high, while only 6.7\% ranked it low. The opposite was the case with missing data experience, with only 13.3\% ranking their experience high and 60\% ranking their experience as low. In addition to the results presented in this section, the descriptive statistics of the results are provided as supplemental material.

\begin{figure}[t]%
       \centering
       \subfloat[][Number of accurate answers.]{
       	\includegraphics[width=8cm]{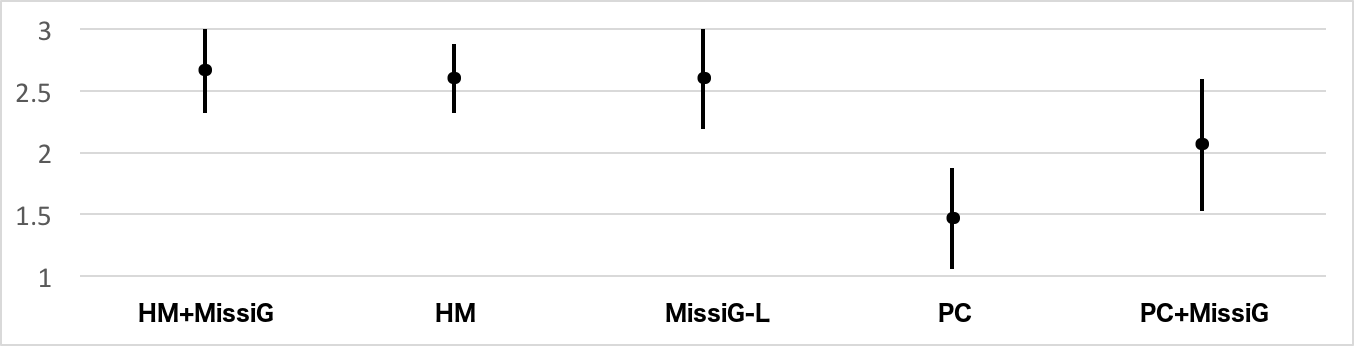}
       	\label{Fig:First_AM_CI_Accuracy}
       }%
       \qquad
       \subfloat[][Average response time in ms.]{
       	\includegraphics[width=8cm]{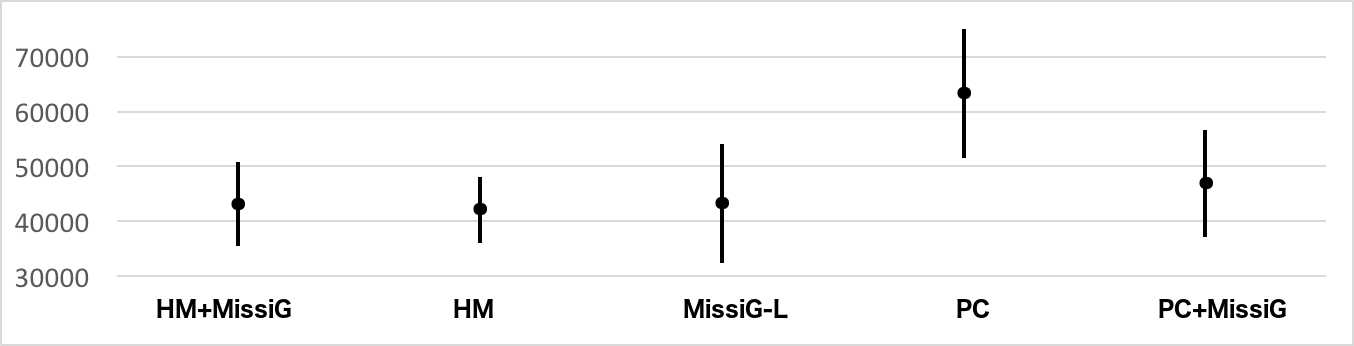}
       	\label{Fig:First_AM_CI_Time}
       	}
       \caption{Confidence intervals for AM tasks}%
       \vspace{-2mm}
       \label{Fig:First_AM_CI}%
\end{figure}

\noindent
\textbf{Amount Missing:} The mean values with 95\% confidence intervals for AM task results are displayed in Fig. \ref{Fig:First_AM_CI}. Statistical testing using Friedman test confirmed significant differences for both accuracy ($\chi^{2}(15)=14.085, p=0.001$) and response time ($\chi^{2}(15)=12.113, p=0.002$) when comparing MissiG with Heatmap and PC. The confidence intervals indicate worse performance for PC both for accuracy and response time. While there is a small overlap in response time between MissiG and PC, research by Cumming and Finch \cite{Cumming2005} conclude that when comparing groups using confidence intervals of individual group estimates, the p-value is near the significance value when the confidence limit of an interval reaches approximately the midpoint between the point estimate and the confidence limit of the other interval. These results were supported by the post-hoc analysis using Wilcoxon signed rank test (table \ref{Table:First_AM_sig}, top) which confirmed significantly worse performance of PC compared to Heatmap and MissiG for AM tasks. For Heatmap compared to HM+MissiG, the confidence intervals overlap and no statistical significance was found for neither accuracy nor response time. PC+MissiG performed better than PC, although the confidence intervals for accuracy overlap. Using Wilcoxon signed rankt test, there was no statistically significant difference for accuracy, while the difference in response time was significant ($Z=-2.669, p=0.008$). These results all support \textbf{H1}, \textbf{H2} and, in part, \textbf{H3}.

\begin{table}[t]
\caption{Pairwise results for MissiG, Heatmap and PC.}\label{Table:First_AM_sig}
	\centering\begin{tabular}{|l|c c| c c|}
\hline
 \textbf{AM} & \multicolumn{2}{c|}{\textbf{Accuracy}} & \multicolumn{2}{c|}{\textbf{Response Time}} \\
Wilcoxon & Z & p & Z & p\\
\hline
HM vs MissiG & 0.000 & 1.000 & -0.284 & 0.776\\
PC vs MissiG & -2.505 & \textcolor{red}{\textbf{0.012}} & -2.726 & \textcolor{red}{\textbf{0.006}}  \\
PC vs HM & -3.017 & \textbf{\textcolor{red}{0.003}} & -3.181 & \textbf{\textcolor{red}{0.001}} \\
\hline
\end{tabular}

\centering\begin{tabular}{|l|c c| c c|}
\hline
 \textbf{JM} & \multicolumn{2}{c|}{\textbf{Accuracy}} & \multicolumn{2}{c|}{\textbf{Response Time}} \\
Wilcoxon & Z & p & Z & p\\
\hline
HM vs MissiG & -1.342 & 0.180 & -2.897 & \textcolor{red}{\textbf{0.004}}\\
PC vs MissiG & -3.373 & \textcolor{red}{\textbf{0.001}} & -3.408 & \textcolor{red}{\textbf{0.001}}  \\
PC vs HM & -3.275 & \textbf{\textcolor{red}{0.001}} & -3.294 & \textbf{\textcolor{red}{0.001}} \\
\hline
\end{tabular}

\centering\begin{tabular}{|l|c c| c c|}
\hline
 \textbf{CM} & \multicolumn{2}{c|}{\textbf{Accuracy}} & \multicolumn{2}{c|}{\textbf{Response Time}} \\
Wilcoxon & Z & p & Z & p\\
\hline
HM vs MissiG & -1.793 & 0.073 & -0.568 & 0.570\\
PC vs MissiG & -1.303 & 0.193 & -3.408 & \textcolor{red}{\textbf{0.001}}  \\
PC vs HM & -.905 & 0.366 & -3.408 & \textbf{\textcolor{red}{0.001}} \\
\hline
\end{tabular}
\end{table}

\begin{figure}[t]%
       \centering
       \subfloat[][Number of accurate answers.]{
       	\includegraphics[width=8cm]{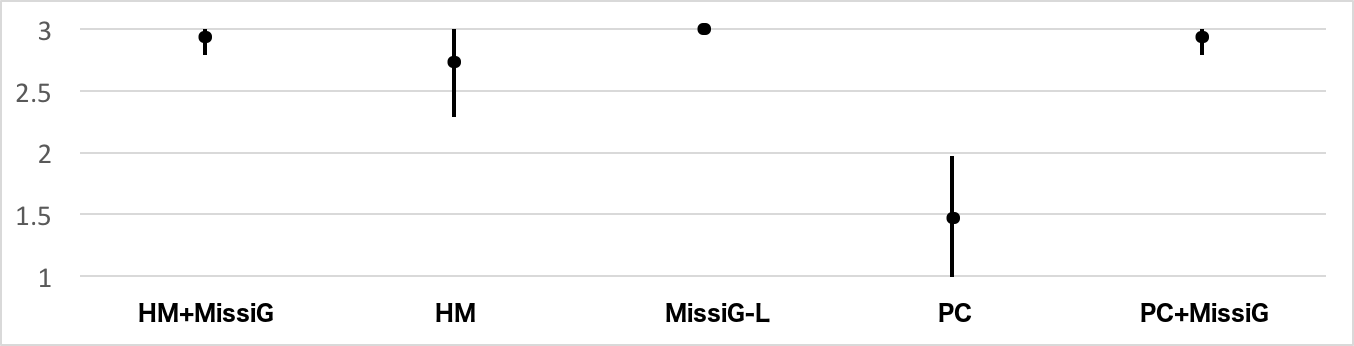}
       	\label{Fig:First_JM_CI_Accuracy}
       }%
       \qquad
       \subfloat[][Average response time in ms.]{
       	\includegraphics[width=8cm]{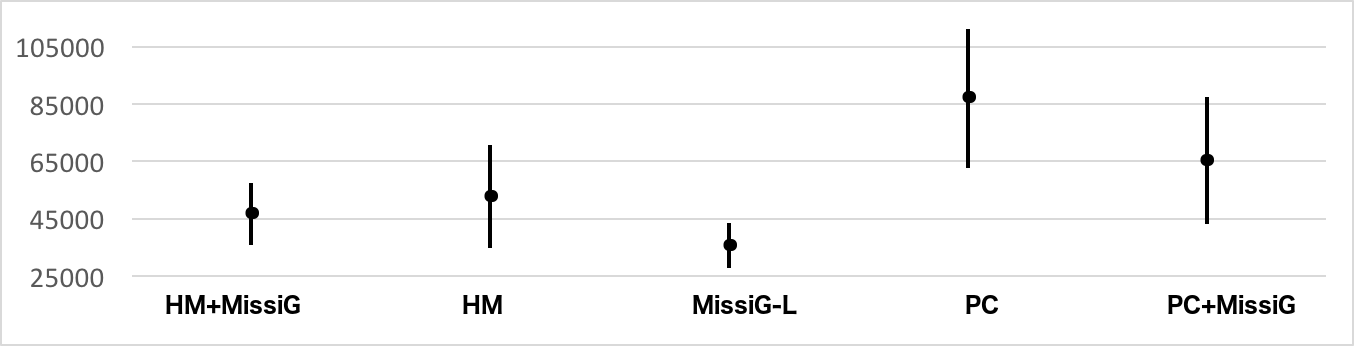}
       	\label{Fig:First_JM_CI_Time}
       	}
       \caption{Confidence intervals for JM tasks}%
       \vspace{-2mm}
       \label{Fig:First_JM_CI}%
\end{figure}

\noindent
\textbf{Joint Missingness:} Fig. \ref{Fig:First_JM_CI} displays the mean values with 95\% confidence intervals for JM task results. Comparing MissiG with Heatmap and PC, the confidence intervals indicate worst performance for PC and slightly better performance for MissiG compared to Heatmap. Friedman test confirm significant differences for both accuracy ($\chi^{2}(15)=25.721, p<0.001$) and response time ($\chi^{2}(15)=24.400, p<0.001$). Post-hoc tests using Wilcoxon signed rank test (table \ref{Table:First_AM_sig}, centre) shows significantly worse result for PC compared to MissiG and Heatmap for both accuracy and response time, and also significantly worse response time for Heatmap compared to MissiG. Although HM+MissiG perform slightly better than Heatmap, the confidence intervals overlap for both accuracy and response time, and no significant differences were found. The performance of PC+MissiG for JM tasks is better than for PC, particularly in terms of accuracy where the two confidence intervals are clearly separated. Analysis using Wilcoxon signed rank test found statistically significant differences for accuracy ($Z=-3.244, p=0.001$) as well as response time ($Z=-2.385, p=0.017$). These results support \textbf{H4}, and, in part, \textbf{H5} and \textbf{H6}.

\begin{figure}[t]%
       \centering
       \subfloat[][Number of accurate answers.]{
       	\includegraphics[width=8cm]{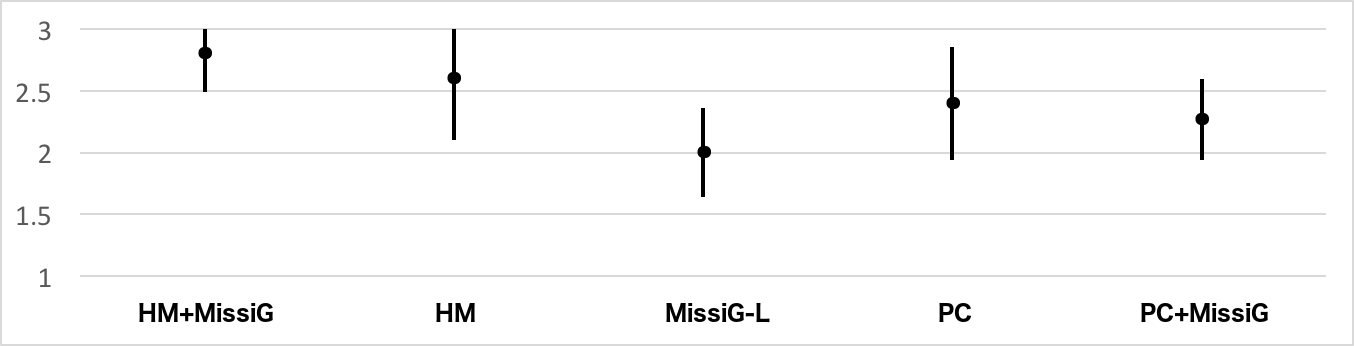}
       	\label{Fig:First_CM_CI_Accuracy}
       }%
       \qquad
       \subfloat[][Average response time in ms.]{
       	\includegraphics[width=8cm]{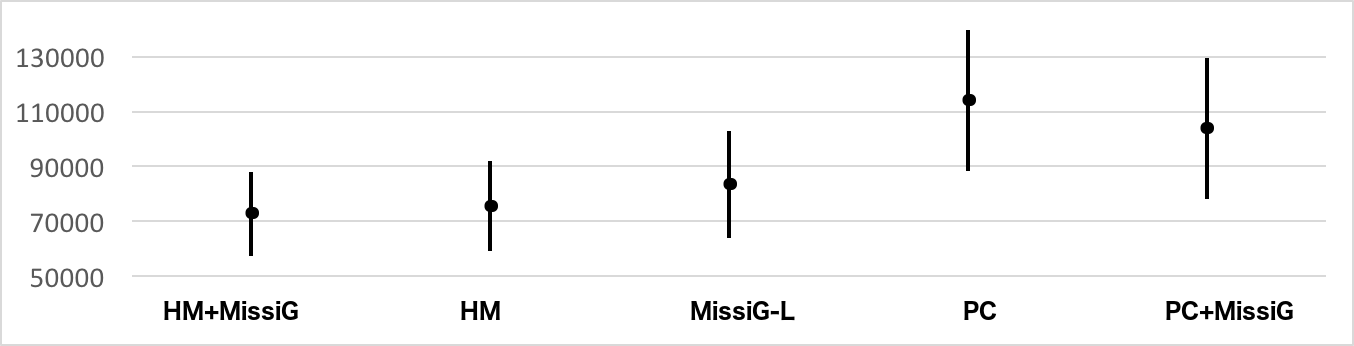}
       	\label{Fig:First_CM_CI_Time}
       	}
       \caption{Confidence intervals for CM tasks}%
       \vspace{-3mm}
       \label{Fig:First_CM_CI}%
\end{figure}

\noindent
\textbf{Conditional Missingness:} The mean values with 95\% confidence intervals for CM task results are displayed in Fig. \ref{Fig:First_CM_CI}. Comparing MissiG with Heatmap and PC the confidence intervals indicate slightly worse accuracy performance for MissiG and worse response time for PC, although the intervals largely overlap for both performance measures. Friedman test results indicate significant differences for both accuracy ($\chi^{2}(15)=6.045, p=0.049$) and response time ($\chi^{2}(15)=22.800, p<0.001$). The Wilcoxon signed rank test (table \ref{Table:First_AM_sig}, bottom) however does not confirm any significant differences for accuracy when the Bonferroni correction has been applied, while the worse response time of PC is significant. The results when comparing HM+MissiG with Heatmap, and PC+MissiG with PC, were not significant. The results for CM tasks hence in part confirms \textbf{H7} (response time for PC is worse than for MissiG), but not \textbf{H8} or \textbf{H9}.

Further to the measured performance, information was gathered with regards to which visualization method the participants found easiest to use, hardest to use and which they preferred to use. A majority, 50\%, found HM+MissiG the easiest visualization to use, followed by 20\% each for MissiG and Heatmap, and 10\% for PC+MissiG. All participants stated that they found PC the hardest to use. The most preferred visualization to use was HM+MissiG, 64.7\%, followed by 11.8\% each for MissiG, Heatmap and PC+MissiG, and 0\% for PC. This generally support \textbf{H10} and indicates that the MissiG design is generally well received by the participants and in particular when combined with Heatmap.

\subsection{Second Study}
The first study was followed by an online study, aiming to further confirm the results and investigate a broader set of tasks related to the three missingness patterns (using all six questions in Section \ref{section:tasks}), as well as including the radial MissiG layout. Furthermore, the second study focused on the visual representation of data, using static images rather than an interactive environment to reduce the potential impact of variation in interactivity across methods. The study compared MissiG (both MissiG-L and MissiG-R) with Heatmap and PC, as well as Heatmap with HM+MissiG, and PC with PC+MissiG. 

\subsubsection{Experimental Design and Procedure}
The experiment was designed as a within-subject study with visualization method as factor. Each participant performed 36 tasks, with one task per question, visualization method and missingness pattern, using the data sets described in \ref{section:data2}. Performance was measured in terms of accuracy and response time. Ethical approval was received prior to the study.

\begin{figure} [t]%
\centering
       	\includegraphics[width=8cm]{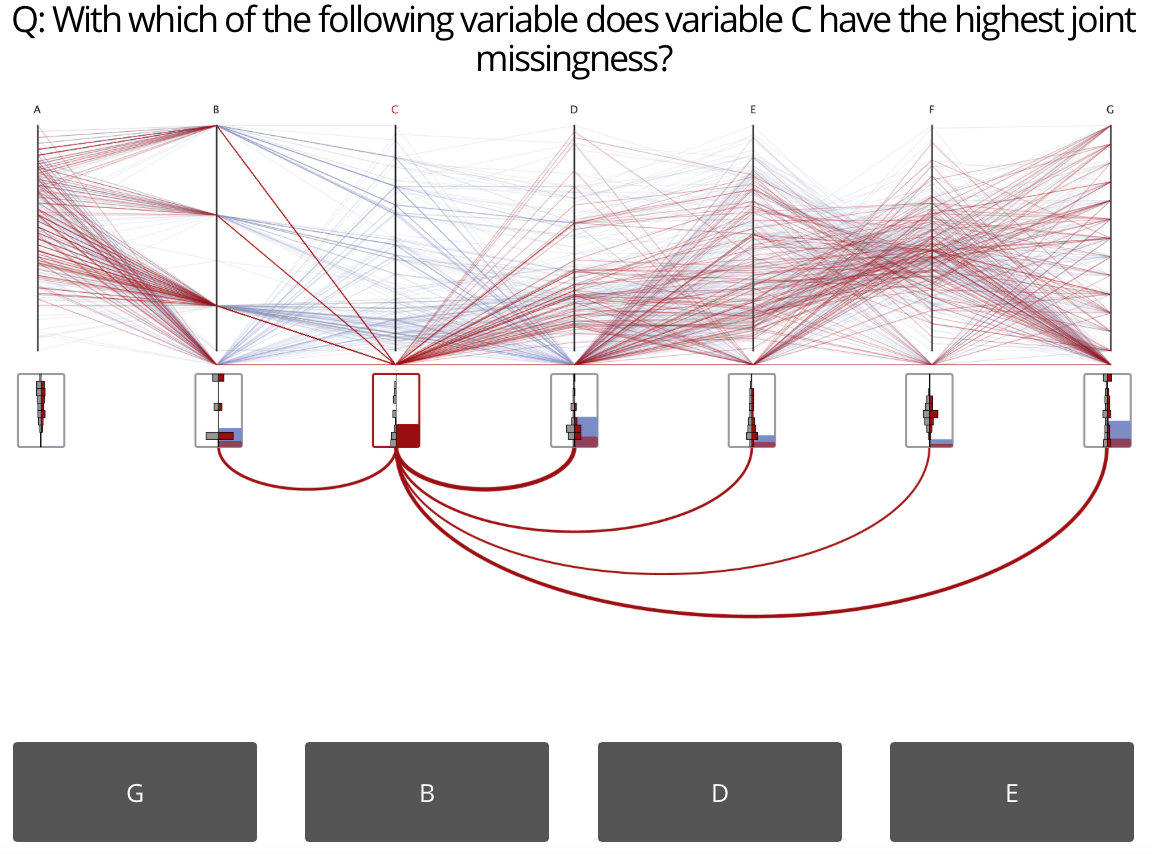}
       \caption{The interface of the second study, with question and visualization at the top and answer buttons at the bottom.}%
       \vspace{-2mm}
       \label{Fig:Study2_interface}%
\end{figure}

The study was conducted online, and implemented using the Gorilla experiment builder \cite{Gorilla}. The experiment was separated into three phases, one for each missingness pattern,  each consisting of a training part and a test part. The training included descriptions of the visualization methods and how to interpret them, in context of the relevant missingness pattern, followed by training using the same type of questions as in the test phase but with feedback on whether the response was accurate to support understanding of the question and visualization. The test phase was similar to the training, with the difference of not including description of visualization and not providing feedback on whether responses were correct or not. The presentation order of visualization methods was randomized for both training and test phases, to reduce the impact of learning effects. The order of missingness patterns were fully counterbalanced, resulting in 6 different orders which were randomly balanced across the participants. The study interface (Fig. \ref{Fig:Study2_interface}) consisted of a question and static image of the visualization method, with a set of multiple choice answers available through buttons. The study was restricted to only run on computers (not mobile phones and tablets) to reduce impact of screen size, and a range of anonymized study data was recorded through Gorilla, of which accuracy and response time was used for the analysis. Background information about participants were collected through a questionnaire, and at the end of each missingness pattern phase the participants were asked to rank their visualization preference for the task.

\subsubsection{Data}\label{section:data2}
As for the first study, a public data set was used and modified through controlled removal of values, to maintain realistic data structures while controlling the missingness patterns in the data. A total of 54 data sets were generated, using the 'cars' \cite{carsdata} data set, which contains real measurements of 392 cars. For each car, seven variables were collected (miles per gallon, number of cylinders, displacement, horsepower, weight, acceleration year, origin). The variable origin that describes where the cars were made (Europe, America and Asia) is categorical and was therefore removed, since some of the visualization methods used (PC in particular) does not perform well with categorical data. The variable names were anonymized and participants did not know which data set was used, thus limiting the impact of preconceptions based on variable names and any need of specific knowledge about cars. 

Missingness patterns were introduced by replacing numerical values with a $NaN$ string. For AM and JM patterns, between 0\% and 40\% of values were randomly removed for each variable. This generated data sets where values are missing completely at random, and JM occur as a result of the random missingness in multiple variables, which works well for the tasks defined in Section \ref{section:tasks} for AM and JM pattern identification. CM patterns required a more controlled removal, with slightly different approaches taken for \textbf{CM1} and \textbf{CM2} tasks. First, between 5\% and 10\% of values were removed randomly from all variables. Then two variables, $X_1$ and $X_2$, were chosen and between 35\% and 70\% of values in $X_1$ were removed for data items with recorded values below the first quartile or above the third quartile in $X_2$. Through this generating CM patterns between missing values in $X_1$ and low or high values in $X_2$. For \textbf{CM2} tasks, which require more than one data set per task, each data set was separated into four subsets with four variables in each, of which one displayed more CM patterns.

\subsubsection{Results}
24 participants initially finished the study. Of these two were removed from analysis due to data quality issues, one of them finishing the whole study in less than three minutes (only possible if not reading instructions and answering without trying to solve tasks), and the other due to the response time to a single question being more than 13 minutes (likely caused by disruption while responding). Of the 22 included participants (5 female, 16 male and 1 preferred not to disclose gender) the majority were between 25 and 44 years old (18-24: 4.5\%, 25-34: 45.5\%, 35-44: 31.8\%, 45-54: 18.2\%, 55 or older: 0\%). Participants were asked to rank their level of experience of: 1) visualization methods, 2) data analysis, and 3) missing data, using 5 point Likert scales ranging from None (1) to Expert (5). 68.2\% ranked their visualization experience as high (4 or 5), while only 9.1\% ranked it as low (1 or 2); 81.8\% ranked their experience of data analysis as high, while none ranked it as low; and 22.7\% ranked their experience of dealing with missing data as high, while 31.8\% ranked it as low.


\begin{figure}[t]%
       \centering
       \subfloat[][Preference of visualization methods for AM tasks.]{
       	\includegraphics[width=8cm]{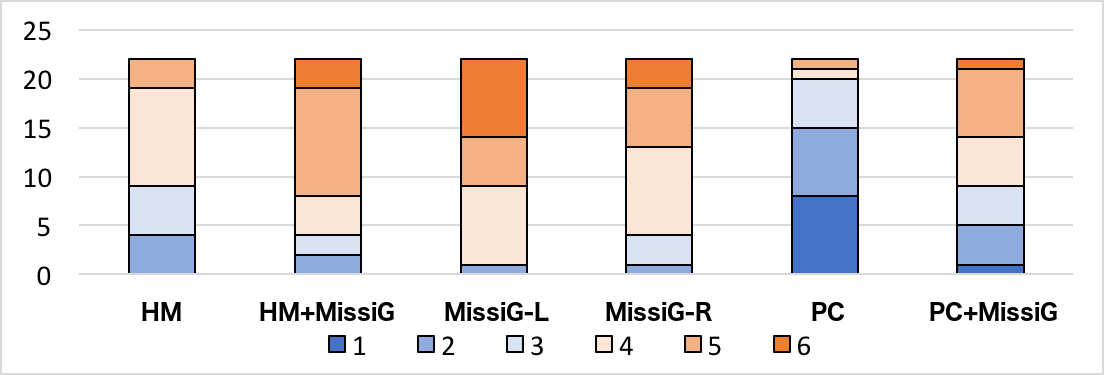}
       	\label{Fig:AM_VisPref}
       }%
       \qquad
       \subfloat[][Preference of visualization methods for JM tasks.]{
       	\includegraphics[width=8cm]{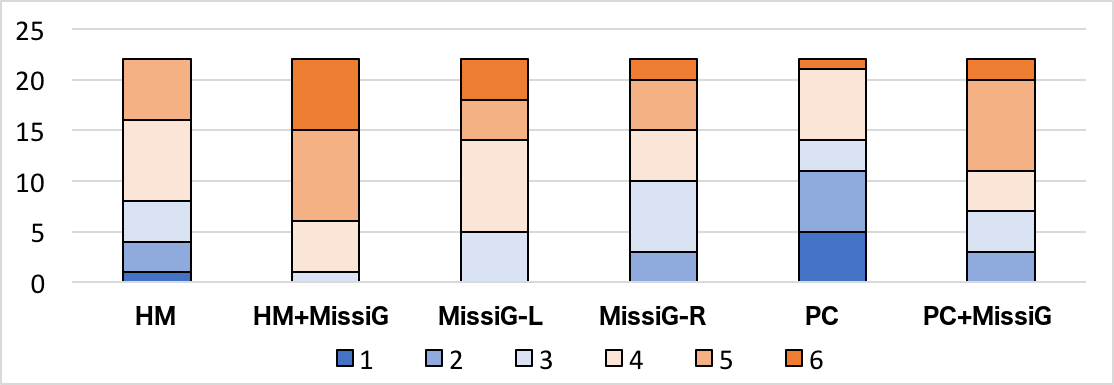}
       	\label{Fig:JM_VisPref}
       	}%
       \qquad
       \subfloat[][Preference of visualization methods for CM tasks. One participant did not provide answer for MissiG-R and PC+MissiG.]{
       	\includegraphics[width=8cm]{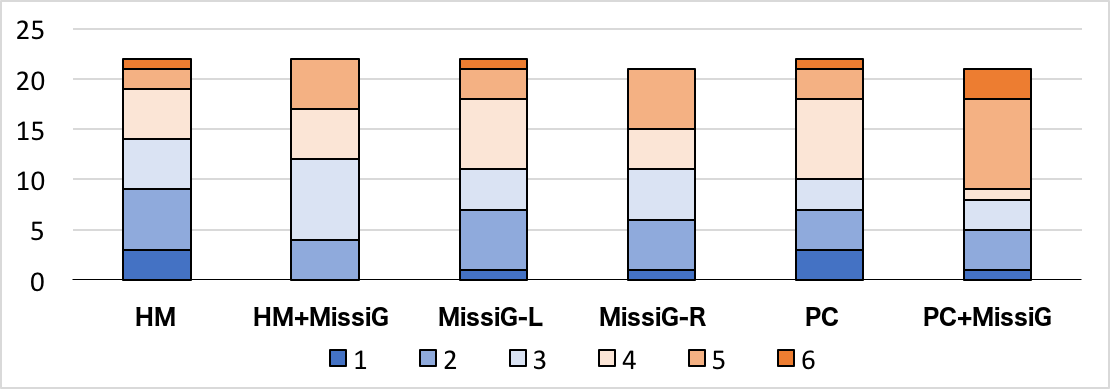}
       	\label{Fig:CM_VisPref}
       }
       \caption{Visualization method preference for different tasks, ranked using a 6 point likert scale ranging from Strongly Disliked (1) to Strongly Liked (6). Blue colour represents negative responses and red represent positive responses.}%
       \vspace{-2mm}
       \label{Fig:VisPreference}%
\end{figure}

\begin{figure}[t]%
       \centering
       \subfloat[][Number of accurate answers.]{
       	\includegraphics[width=8cm]{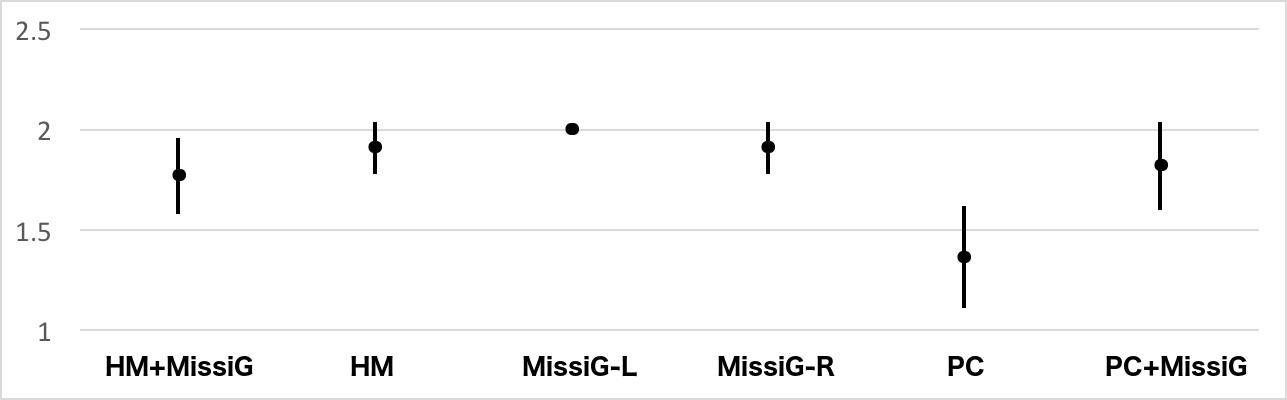}
       	\label{Fig:AM_CI_AccResult}
       }%
       \qquad
       \subfloat[][Average response time in ms.]{
       	\includegraphics[width=8cm]{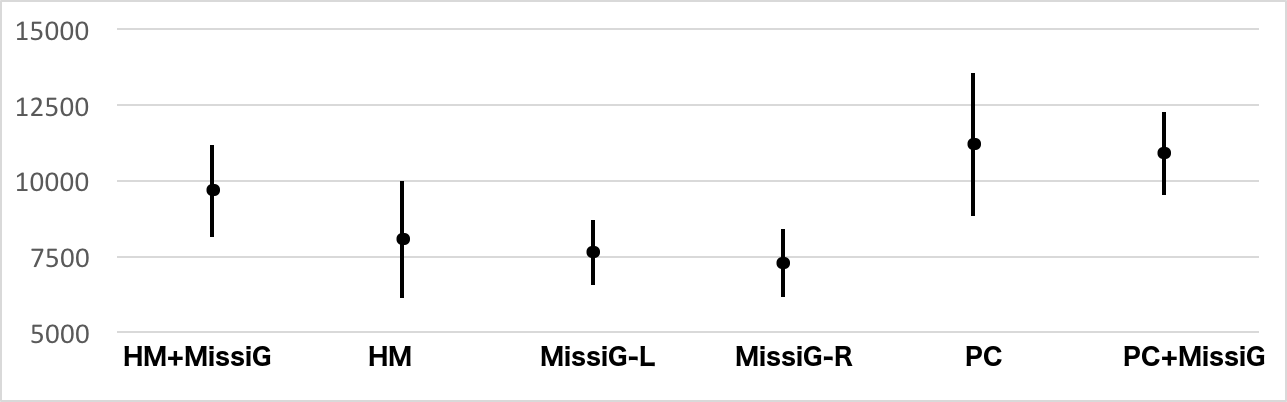}
       	\label{Fig:AM_CI_TimeRes}
       	}%
       \qquad
       \subfloat[][Response time in ms for correct answers only.]{
       	\includegraphics[width=8cm]{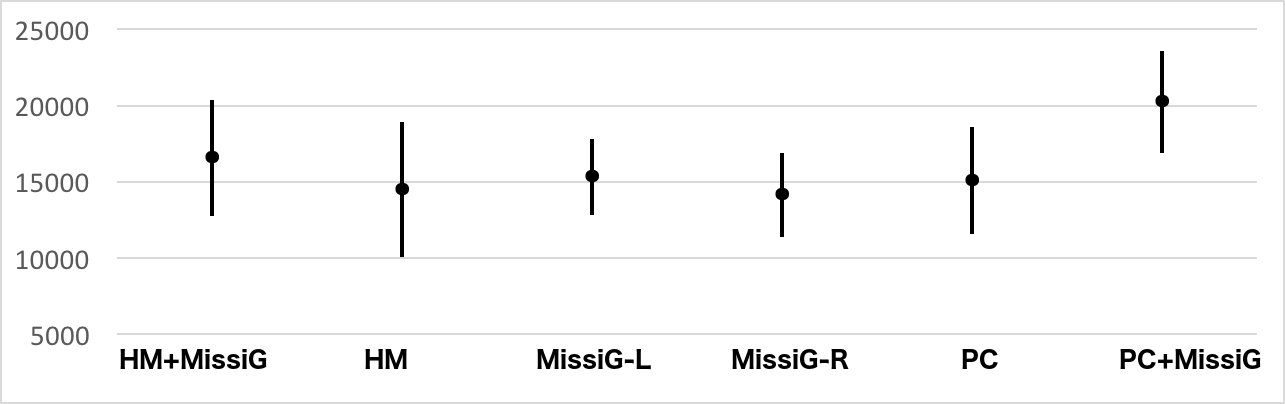}
       	\label{Fig:AM_CI_CorrTimeRes}
       }
       \caption{Confidence intervals for AM tasks}%
       \vspace{-2mm}
       \label{Fig:AM_ConfidenceIntervals}%
\end{figure}

\noindent
\textbf{Amount Missing:} Fig. \ref{Fig:AM_VisPref} displays a summary of the preference of the participants for each visualization method for AM tasks, using a likert scale ranging from 1 (strongly disliked) to 6 (strongly liked). From the figure we can conclude that MissiG-L was the most preferred visualization method for AM tasks, followed by HM+MissiG and MissiG-R, which supports \textbf{H10}. PC was the least liked method for AM tasks. Confidence intervals for the performance of the visualization methods are presented in Fig. \ref{Fig:AM_ConfidenceIntervals}. When comparing the two MissiG layouts with Heatmap and PC, Friedman tests confirmed statistically significant differences in accuracy ($\chi^{2}(22)=29.110, p<0.001$) and response time ($\chi^{2}(22)=20.018, p<0.001$), while differences for response time only for accurate answers were not significant. For accuracy and response time, Wilcoxon signed rank test (table \ref{AM_table}) revealed significantly worse performance for PC compared to all other visualization methods. No other differences were significant. These results confirm that MissiG performs better than PC, and equally good or better than Heatmap for AM tasks (\textbf{H1}). Wilcoxon tests revealed significant difference in response time for Heatmap compared to HM+MissiG ($Z=-2.127, p=0.033$) with better performance for Heatmap, while differences in accuracy and response time for correct answers were not significant. These results does not confirm \textbf{H3}. For PC compared to PC+MissiG, Wilcoxon tests showed significant differences for accuracy ($Z=-2.202, p=0.028$) and response time for correct answers ($Z=-2.315, p=0.021$), with higher accuracy but slower response time when PC is enhanced with MissiG. The overall difference in response time was however not significant. This support \textbf{H2} for accuracy, but not response time. An explanation to slower response times for visualization methods enhanced with MissiG can be the additional cognitive burden of combining two visualization methods instead of one, as well as the likely unfamiliarity of the MissiG visualization. The accuracy results, which generally would be more important than response time, however does support the benefit of enhancing PC with MissiG.

\begin{table}[t]
\caption{Pairwise results for AM tasks for MissiG-L, MissiG-R, Heatmap (HM) and PC.}\label{AM_table}
	\centering\begin{tabular}{|l|c c| c c|}
\hline
 & \multicolumn{2}{c|}{\textbf{Accuracy}} & \multicolumn{2}{c|}{\textbf{Response Time}} \\
Wilcoxon & Z & p & Z & p\\
\hline
MissiG-L vs HM & -1.414 & 0.157 & -0.081 & 0.935\\
MissiG-R vs HM & 0.000 & 1.000 & -1.088 & 0.277 \\
PC vs HM & -3.207 & \textcolor{red}{\textbf{0.001}} & -2.646 & \textcolor{red}{\textbf{0.008}}  \\
MissiG-R vs MissiG-L & -1.414 & 0.157 & -0.828 & 0.408\\
PC vs MissiG-L & -3.500 & \textbf{\textcolor{red}{$<$0.001}} & -3.685 & \textbf{\textcolor{red}{$<$0.001}} \\
PC vs MissiG-R & -3.207 & \textcolor{red}{\textbf{0.001}} & -3.393 & \textbf{\textcolor{red}{0.001}}\\
\hline
\end{tabular}
\end{table}

\begin{figure}[t]%
       \centering
       \subfloat[][Number of accurate answers.]{
       	\includegraphics[width=8cm]{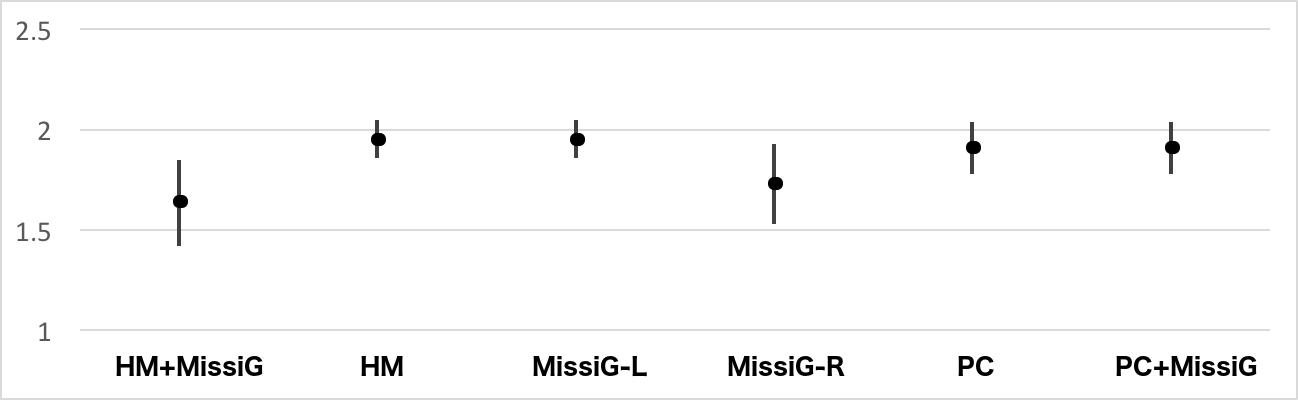}
       	\label{Fig:JM_CI_AccResult}
       }%
       \qquad
       \subfloat[][Average response time in ms.]{
       	\includegraphics[width=8cm]{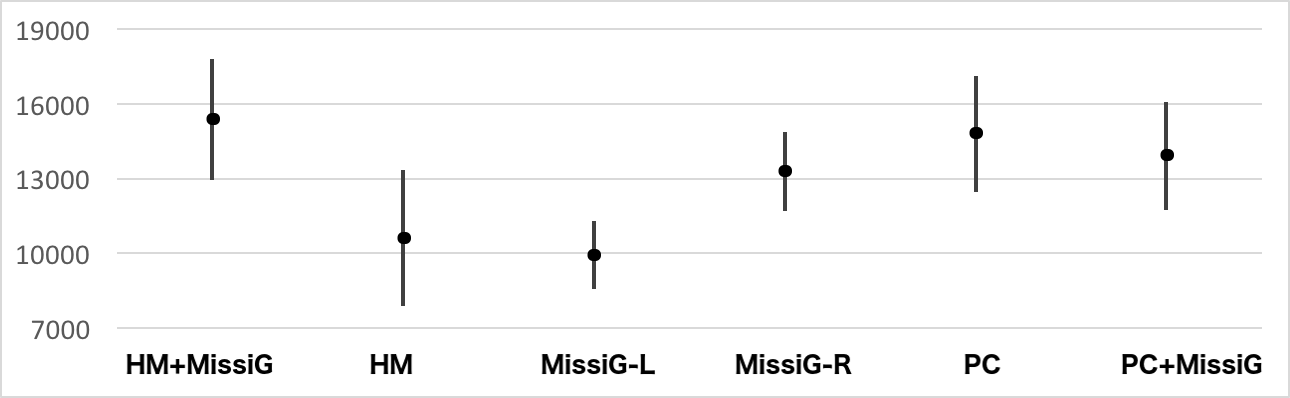}
       	\label{Fig:JM_CI_TimeRes}
       	}%
       \qquad
       \subfloat[][Response time in ms for correct answers only.]{
       	\includegraphics[width=8cm]{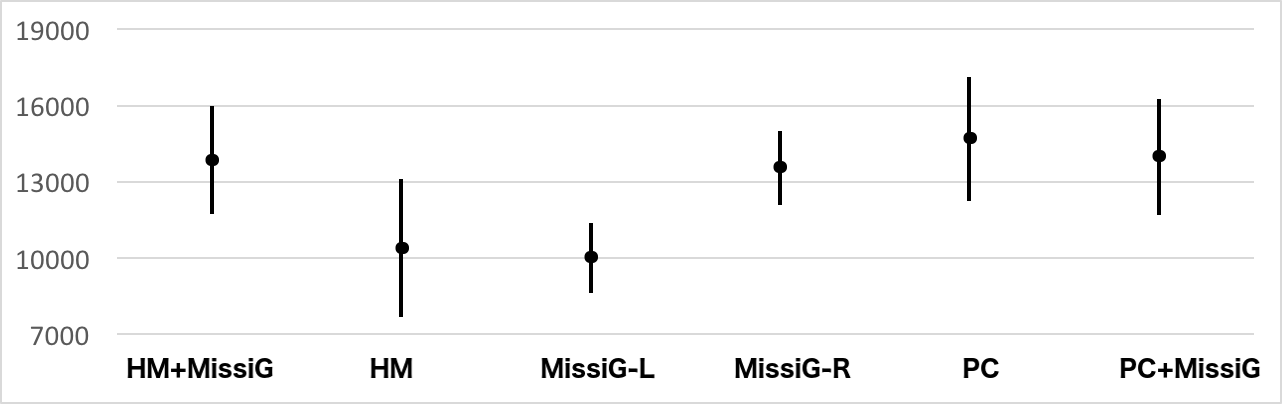}
       	\label{Fig:JM_CI_CorrTimeRes}
       }
       \caption{Confidence intervals for JM tasks}%
       \vspace{-1.5mm}
       \label{Fig:JM_ConfidenceIntervals}%
\end{figure}

\noindent
\textbf{Joint Missingness:} Fig. \ref{Fig:JM_VisPref} displays a summary of the preference of the participants for each visualization method for JM tasks. From the figure we can conclude that HM+MissiG was the most preferred visualization method for JM tasks, followed by MissiG-L, PC+MissiG and Heatmap, which generally confirms \textbf{H10} for JM tasks. PC was the least liked method also for JM tasks. Confidence intervals for the performance of the visualization methods for JM tasks are presented in Fig. \ref{Fig:JM_ConfidenceIntervals}. When comparing the two MissiG layouts with Heatmap and PC, Friedman tests confirmed statistically significant differences in response time ($\chi^{2}(22)=32.073, p<0.001$) and response time only for accurate answers ($\chi^{2}(22)=30.491, p<0.001$), while the differences for accuracy were not significant. For response time and response time of accurate answers, Wilcoxon signed rank post-hoc tests (table \ref{Table:JM_sig}) revealed significantly better performance for Heatmap and MissiG-L compared to MissiG-R and PC, which support \textbf{H4} for response time, although not for accuracy. Wilcoxon tests revealed that the difference in performance for Heatmap compared to HM+MissiG was significant for accuracy ($Z=-2.646, p=0.008$), response time ($Z=-3.360, p=0.001$) and response time only for accurate answers ($Z=-3.295, p=0.001$), with overall better performance for Heatmap. There were no significant performance differences for PC compared to PC+MissiG. Thus, \textbf{H5} and \textbf{H6} are not supported by these results.

\begin{table}[t]
\caption{Pairwise results for JM tasks for MissiG-L, MissiG-R, Heatmap (HM) and PC.}\label{Table:JM_sig}
	\centering\begin{tabular}{|l|c c|}
\hline
 & \multicolumn{2}{c|}{\textbf{Response Time}} \\
Wilcoxon & Z & p \\
\hline
MissiG-L vs HM & -0.276 & 0.783 \\
MissiG-R vs HM & -3.133 & \textcolor{red}{\textbf{0.002}}   \\
PC vs HM & -3.328 & \textcolor{red}{\textbf{0.001}}  \\
MissiG-R vs MissiG-L & -3.782 & \textbf{\textcolor{red}{$<$0.001}} \\
PC vs MissiG-L & -3.782 & \textbf{\textcolor{red}{$<$0.001}}  \\
PC vs MissiG-R & -0.828 & 0.408 \\
\hline
 & \multicolumn{2}{c|}{\textbf{Response Time Correct}} \\
Wilcoxon & Z & p \\
\hline
MissiG-L vs HM & -0.893 & 0.372 \\
MissiG-R vs HM & -3.198 & \textcolor{red}{\textbf{0.001}}   \\
PC vs HM & -3.100 & \textcolor{red}{\textbf{0.002}}  \\
MissiG-R vs MissiG-L & -3.912 & \textbf{\textcolor{red}{$<$0.001}} \\
PC vs MissiG-L & -3.263 & \textcolor{red}{\textbf{0.001}}  \\
PC vs MissiG-R & -0.990 & 0.322 \\
\hline
\end{tabular}
\vspace{-1mm}
\end{table}

\begin{figure}[t]%
       \centering
       \subfloat[][Number of accurate answers.]{
       	\includegraphics[width=7.9cm]{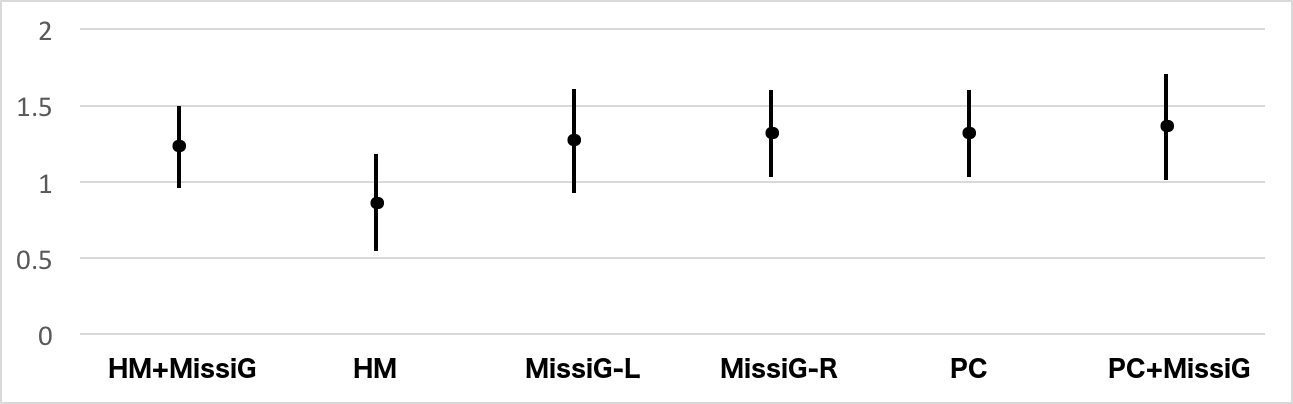}
       	\label{Fig:CM_CI_AccResult}
       }%
       \qquad
       \subfloat[][Average response time in ms.]{
       	\includegraphics[width=7.9cm]{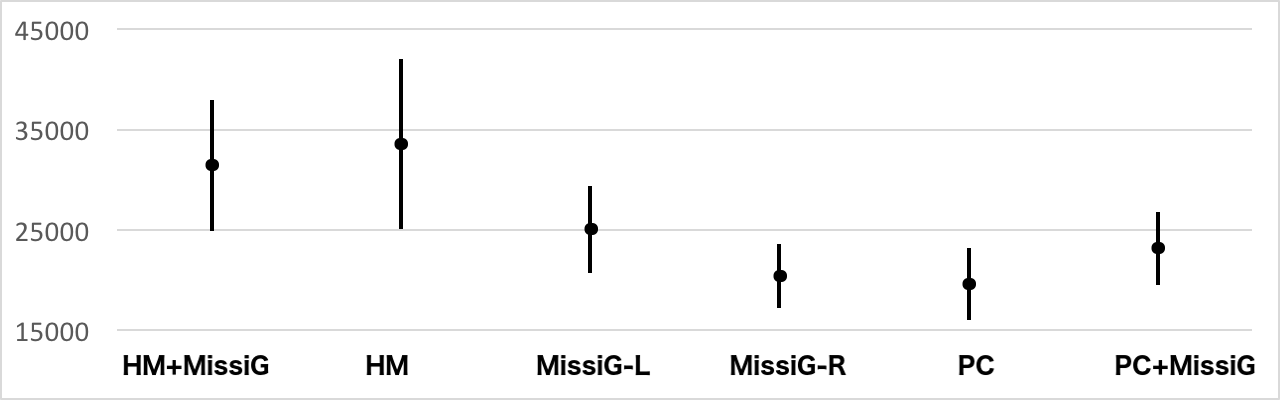}
       	\label{Fig:CM_CI_TimeRes}
       	}%
       \qquad
       \subfloat[][Response time in ms for correct answers only.]{
       	\includegraphics[width=7.9cm]{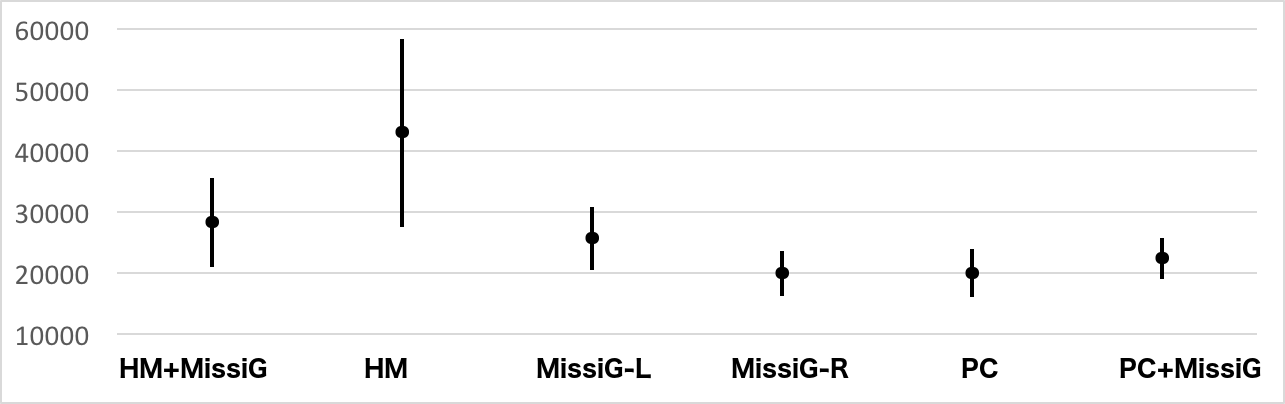}
       	\label{Fig:CM_CI_CorrTimeRes}
       }
       \caption{Confidence intervals for CM tasks}%
       \vspace{-1mm}
       \label{Fig:CM_ConfidenceIntervals}%
\end{figure}

\begin{table}[t]
\caption{Pairwise results for CM tasks for MissiG-L, MissiG-R, Heatmap (HM) and PC.}\label{Table:CM_sig}
	\centering\begin{tabular}{|l|c c|}
\hline
 & \multicolumn{2}{c|}{\textbf{Response Time}} \\
ANOVA &  & p \\
\hline
MissiG-L vs HM &  & 0.135 \\
MissiG-R vs HM &  & \textcolor{red}{\textbf{0.006}}   \\
PC vs HM &  & \textcolor{red}{\textbf{0.004}}  \\
MissiG-R vs MissiG-L &  & 0.033 \\
PC vs MissiG-L &  & \textcolor{red}{\textbf{0.005}}  \\
PC vs MissiG-R &  & 1.000 \\
\hline
 & \multicolumn{2}{c|}{\textbf{Response Time Correct}} \\
Wilcoxon & Z & p  \\
\hline
MissiG-L vs HM & -2.731 & \textcolor{red}{\textbf{0.006}} \\
MissiG-R vs HM & -2.953 & \textcolor{red}{\textbf{0.003}}   \\
PC vs HM & -2.897 & \textcolor{red}{\textbf{0.004}}  \\
MissiG-R vs MissiG-L & -2.959 & \textcolor{red}{\textbf{0.003}} \\
PC vs MissiG-L & -3.243 & \textcolor{red}{\textbf{0.001}}  \\
PC vs MissiG-R & -0.121 & 0.904 \\
\hline
\end{tabular}
\vspace{-1mm}
\end{table}

\noindent
\textbf{Conditional Missingness:} Fig. \ref{Fig:CM_VisPref} displays a summary of the preference of the participants for each visualization method for CM tasks. It can be concluded that PC+MissiG was the most preferred visualization method for CM tasks, followed by PC, MissiG-L, and MissiG-R, again confirming that the participants liked using MissiG as part of the analysis (\textbf{H10}). Heatmap was the least liked method. Confidence intervals for the performance of the visualization methods are presented in Fig. \ref{Fig:CM_ConfidenceIntervals}. When comparing the two MissiG layouts with Heatmap and PC, tests confirmed statistically significant difference in response time (ANOVA: $F(1.452, 30.499)=11.906, p=0.001$) and response time for accurate answers  (Friedman: $\chi^{2}(14)=20.229, p<0.001$), but not for accuracy. Post-hoc analysis with ANOVA and Wilcoxon signed rank test (table \ref{Table:CM_sig}) showed that MissiG-R and PC performed significantly better than Heatmap for both response time and response time for accurate answers, and PC performed significantly better than MissiG-L for both response times as well, which is also clear from the confidence intervals. Additionally, MissiG-L performed significantly better than Heatmap, and MissiG-R performed significantly better than MissiG-L for response time for accurate answers. This supports \textbf{H7} in terms of response time, while accuracy results are inconclusive. Looking at the results in Fig. \ref{Fig:CM_ConfidenceIntervals}, HM+MissiG generally performs better than Heatmap, which supports \textbf{H9}, although Wilcoxon signed rank test show that the difference is only significant for response time for correct answers ($Z=-2.731, p=0.006$). A  dependent t-Test showed that the difference in response time for PC compared to PC+MissiG, with slightly better performance for PC, was significant ($t(21)=-2.2374, p=0.027$), which does not support \textbf{H8}, while there was no difference for accuracy or response time for accurate answers.



\section{Discussion and Conclusions}\label{section:conclusions}
This paper presented MissiG, a novel glyph based method for visualization of missing values and missingness patterns in data. The method was designed based on established glyph. design guidelines to focus on the representation of three patterns of importance for understanding missingness in data, namely the amount missing in variables, the joint missingness in pairs of variables, and the conditional missingness between missing and recorded values in variables. MissiG can be used both as a standalone multivariate visualization method, with two different layouts presented in the paper, and as an enhancement to existing visualization methods, here demonstrated through enhancement of Heatmap and PC. These four methods were evaluated against Heatmap and PC through two usability studies. 

The results from the studies indicate that MissiG performs better than PC and equally well as Heatmap for amount missing and joint missingness tasks. The results for conditional missingness tasks indicate that MissiG is equally good as PC and may be better than Heatmap, although the results are not conclusive across the two studies. This may be due to the concept of conditional missingness being more complex than amount missing and joint missingness, which was indicated by questions made by participants in the first study, and thus can have resulted in a higher degree of uncertainty in the results. It is also worth noting that the lack of difference for accuracy in some cases may be the result of ceiling effects when a majority of participants answered correctly to most questions. The difference in results across the two evaluations can furthermore be a result of the different study designs, with the first study using interactive visualization and the second study including additional questions. This difference was intentional to cover a broader range of analysis situations and tasks, but may have impacted results. In terms of using MissiG as an enhancement, PC with MissiG generally performed better than PC for amount missing in both studies and joint missingness in the first study, while it did not perform better for conditional missingness tasks. Conversely, Heatmap enhanced with MissiG did not perform as well as Heatmap for amount missing and joint missingness tasks, but did in part perform better for conditional missingness, although results were not conclusive across the two studies. Several aspects may have influenced these results. Firstly, the analysis of two coordinated views compared to a single view may have a higher cognitive burden, resulting in longer response times for the enhanced visualization methods. Furthermore, the majority of participants ranked their experience of visualization as high rather than low, which suggests that they may have had previous experience of Heatmap and PC, whereas MissiG is a novel visualization method. This can have impacted MissiG negatively, particularly in terms of response time. It is worth noting that enhancement with MissiG seems to be mostly beneficial for tasks where the basic visualization method is limited (i.e. amount missing and joint missingness for PC, and conditional missingness for Heatmap), whereas the benefit of enhancement is more questionable for tasks where the basic method already performs well. The studies confirmed that MissiG is the preferred choice by users as a method for visualizing missingness in data, which further confirms its usability and the potential of its utility as part of more complex visual analysis workflows. To summarize the main results:
\begin{itemize}
    \item MissiG performs better than PC, and equally well as Heatmap, for amount missing and joint missingness.
    \item MissiG in part performs better than Heatmap, and equally well as PC, for conditional missingness.
    \item PC+MissiG generally performs better than PC for amount missing and joint missingness.
    \item Heatmap+MissiG performs equally well and in part better than Heatmap for conditional missingness.
    \item Visualization with MissiG was generally the preferred choice by users across all three missingness patterns.
\end{itemize}

These results are encouraging and strongly suggest that MissiG has potential to greatly improve analysis and understanding of missingness patterns in data, and through this support decision making as to how to deal with missing values, as well as to reveal important insights related to missing values. Future work includes the application and qualitative testing of MissiG with application domain experts in substantially more complex analysis settings. Of interest would be to apply the technique to IoT and sensor data. In domains such as Air traffic control and unmanned aerial management, as well as in medical digital mobility assessment, where missing values may have a pertinent meaning and where many parameters interplay with each other. Using this kind of tool to analyse relationships between missing data could potentially give a better understanding of how to improve such systems. Utilizing the flexibility of the glyph design to provide additional layout options and enhancement of more visualization methods is a topic for immediate future work.

\ifCLASSOPTIONcaptionsoff
  \newpage
\fi


\begin{balance}
\bibliographystyle{IEEEtran}
\bibliography{refs}

\begin{thebibliography}{10}
\providecommand{\url}[1]{#1}
\csname url@samestyle\endcsname
\providecommand{\newblock}{\relax}
\providecommand{\bibinfo}[2]{#2}
\providecommand{\BIBentrySTDinterwordspacing}{\spaceskip=0pt\relax}
\providecommand{\BIBentryALTinterwordstretchfactor}{4}
\providecommand{\BIBentryALTinterwordspacing}{\spaceskip=\fontdimen2\font plus
\BIBentryALTinterwordstretchfactor\fontdimen3\font minus
  \fontdimen4\font\relax}
\providecommand{\BIBforeignlanguage}[2]{{%
\expandafter\ifx\csname l@#1\endcsname\relax
\typeout{** WARNING: IEEEtran.bst: No hyphenation pattern has been}%
\typeout{** loaded for the language `#1'. Using the pattern for}%
\typeout{** the default language instead.}%
\else
\language=\csname l@#1\endcsname
\fi
#2}}
\providecommand{\BIBdecl}{\relax}
\BIBdecl

\bibitem{Carpenter2013}
J.~Carpenter and M.~Kenward, \emph{Multiple Imputation and its
  Application}.\hskip 1em plus 0.5em minus 0.4em\relax Wiley, 2013.

\bibitem{Fernstad2019}
S.~{Johansson Fernstad}, ``To identify what isn't there: A definition of
  missingness patterns and evaluation of missing value visualization,''
  \emph{Information Visualization}, vol.~18, no.~2, pp. 230--250, 2019.

\bibitem{Fernstad2014}
S.~{Johansson Fernstad} and R.~C. Glen, ``Visual analysis of missing data -- to
  see what isn't there,'' in \emph{Poster Proceedings of IEEE Vis}.\hskip 1em
  plus 0.5em minus 0.4em\relax IEEE, November 2014.

\bibitem{Chung2015}
D.~H. Chung, P.~A. Legg, M.~L. Parry, R.~Bown, I.~W. Griffiths, R.~S. Laramee,
  and M.~Chen, ``Glyph sorting: Interactive visualization for multi-dimensional
  data,'' \emph{Information Visualization}, vol.~14, no.~1, pp. 76--90, 2015.

\bibitem{Maguire2012}
E.~Maguire, P.~Rocca-Serra, S.-A. Sansone, J.~Davies, and M.~Chen,
  ``Taxonomy-based glyph design—with a case study on visualizing workflows of
  biological experiments,'' \emph{IEEE Transactions on Visualization and
  Computer Graphics}, vol.~18, no.~12, pp. 2603--2612, 2012.

\bibitem{borgo2013}
R.~Borgo, J.~Kehrer, D.~H. Chung, E.~Maguire, R.~S. Laramee, H.~Hauser,
  M.~Ward, and M.~Chen, ``Glyph-based visualization: Foundations, design
  guidelines, techniques and applications.'' in \emph{Eurographics (STARs)},
  2013, pp. 39--63.

\bibitem{Rubin1976}
D.~B. Rubin, ``Inference and missing data,'' \emph{Biometrika}, vol.~63, no.~3,
  pp. 581--592, 1976.

\bibitem{Wang2007}
H.~Wang and S.~Wang, ``Visualization of the critical patterns of missing values
  in classification data,'' in \emph{International Conference on Advances in
  Visual Information Systems}.\hskip 1em plus 0.5em minus 0.4em\relax Springer,
  2007, pp. 267--274.

\bibitem{Wang2009}
H.~Wang and S.~{Wa}ng, ``Data mining with incomplete data,'' in \emph{Data
  Warehousing and Mining: Concepts, Methodologies, Tools, and
  Applications}.\hskip 1em plus 0.5em minus 0.4em\relax IGI Global, 2008, pp.
  3027--3032.

\bibitem{Fielding2009}
S.~Fielding, P.~M. Fayers, and C.~R. Ramsay, ``Investigating the missing data
  mechanism in quality of life outcomes: a comparison of approaches,''
  \emph{Health and Quality of Life Outcomes}, vol.~7, no.~1, p.~57, 2009.

\bibitem{Djurcilov2000}
S.~Djurcilov and A.~Pang, ``Visualizing sparse gridded data sets,'' \emph{IEEE
  Computer Graphics and Applications}, vol.~20, no.~5, pp. 52--57, 2000.

\bibitem{Beddow1990}
J.~Beddow, ``Shape coding of multidimensional data on a microcomputer
  display,'' in \emph{Proceedings of the 1st Conference on
  Visualization'90}.\hskip 1em plus 0.5em minus 0.4em\relax IEEE Computer
  Society Press, 1990, pp. 238--246.

\bibitem{Twiddy1994}
R.~Twiddy, J.~Cavallo, and S.~M. Shiri, ``Restorer: A visualization technique
  for handling missing data,'' in \emph{Proceedings of the conference on
  Visualization'94}.\hskip 1em plus 0.5em minus 0.4em\relax IEEE Computer
  Society Press, 1994, pp. 212--216.

\bibitem{Unwin1996}
A.~Unwin, G.~Hawkins, H.~Hofmann, and B.~Siegl, ``Interactive graphics for data
  sets with missing values — manet,'' \emph{Journal of Computational and
  Graphical Statistics}, vol.~5, no.~2, pp. 113--122, 1996.

\bibitem{Theus1997}
M.~Theus, H.~Hofmann, B.~Siegl, and A.~Unwin, ``Manet extensions to interactive
  statistical graphics for missing values,'' in \emph{In New Techniques and
  Technologies for Statistics II}.\hskip 1em plus 0.5em minus 0.4em\relax IOS
  Press, 1997, pp. 247--259.

\bibitem{Swayne1998}
D.~F. Swayne and A.~Buja, ``Missing data in interactive high-dimensional data
  visualization,'' \emph{Computational Statistics}, vol.~13, no.~1, pp. 15--26,
  1998.

\bibitem{Swayne2003}
D.~F. Swayne, D.~T. Lang, A.~Buja, and D.~Cook, ``Ggobi: Evolving from xgobi
  into an extensible framework for interactive data visualization,''
  \emph{Comput. Stat. Data Anal.}, vol.~43, no.~4, pp. 423--444, Aug. 2003.

\bibitem{Kohonen1998}
T.~Kohonen, ``The self-organizing map,'' \emph{Neurocomputing}, vol.~21, no.
  1--3, pp. 1--6, 1998.

\bibitem{unwin2015graphical}
A.~Unwin, \emph{Graphical data analysis with R}.\hskip 1em plus 0.5em minus
  0.4em\relax CRC Press, 2015, vol.~27.

\bibitem{tierney2019naniar}
N.~Tierney, D.~Cook, M.~McBain, C.~Fay, M.~O'Hara-Wild, J.~Hester, and
  L.~Smith, ``Naniar: Data structures, summaries, and visualizations for
  missing data,'' \emph{R Package}, 2019.

\bibitem{lex2014upset}
A.~Lex, N.~Gehlenborg, H.~Strobelt, R.~Vuillemot, and H.~Pfister, ``Upset:
  visualization of intersecting sets,'' \emph{IEEE transactions on
  visualization and computer graphics}, vol.~20, no.~12, pp. 1983--1992, 2014.

\bibitem{pilhoefer2014extracat}
A.~Pilhoefer, ``Extracat: Categorical data analysis and visualization,''
  \emph{R package version}, pp. 1--7, 2014.

\bibitem{Templ2012}
M.~Templ, A.~Alfons, and P.~Filzmoser, ``Exploring incomplete data using
  visualization techniques,'' \emph{Advances in Data Analysis and
  Classification}, vol.~6, no.~1, pp. 29--47, 2012.

\bibitem{Brix2011}
P.~Brix, ``mi{P}: Multiple imputation plots,'' https://CRAN.R-project.org/
  package=miP, 2012.

\bibitem{Cheng2015}
X.~Cheng, D.~Cook, H.~Hofmann \emph{et~al.}, ``Visually exploring missing
  values in multivariable data using a graphical user interface,''
  \emph{Journal of Statistical Software}, vol.~68, no.~6, pp. 1--23, 2015.

\bibitem{Honaker2011}
J.~Honaker, G.~King, and M.~Blackwell, ``Amelia ii: A program for missing
  data,'' \emph{Journal of Statistical Software, Articles}, vol.~45, no.~7, pp.
  1--47, 2011.

\bibitem{kandel2012}
S.~Kandel, R.~Parikh, A.~Paepcke, J.~M. Hellerstein, and J.~Heer, ``Profiler:
  Integrated statistical analysis and visualization for data quality
  assessment,'' in \emph{Proceedings of the International Working Conference on
  Advanced Visual Interfaces}, 2012, pp. 547--554.

\bibitem{Gschwandtner2018}
T.~Gschwandtner and O.~Erhart, ``Know your enemy: Identifying quality problems
  of time series data,'' in \emph{2018 IEEE Pacific Visualization Symposium
  (PacificVis)}.\hskip 1em plus 0.5em minus 0.4em\relax IEEE, 2018, pp.
  205--214.

\bibitem{Triana2019}
J.~A. Triana, D.~Zeckzer, H.~Hagen, and J.~T. Hernandez, ``Vafusq: A
  methodology to build visual analysis applications with data quality
  features,'' \emph{Information Visualization}, vol.~18, no.~4, pp. 384--404,
  2019.

\bibitem{Schulz2017}
H.-J. Schulz, T.~Nocke, M.~Heitzler, and H.~Schumann, ``A systematic view on
  data descriptors for the visual analysis of tabular data,'' \emph{Information
  Visualization}, vol.~16, no.~3, pp. 232--256, 2017.

\bibitem{Cedilnik2000}
A.~Cedilnik and P.~Rheingans, ``Procedural annotation of uncertain
  information,'' in \emph{Visualization 2000. Proceedings}.\hskip 1em plus
  0.5em minus 0.4em\relax IEEE, 2000, pp. 77--84.

\bibitem{Xie2006}
Z.~Xie, S.~Huang, M.~O. Ward, and E.~A. Rundensteiner, ``Exploratory
  visualization of multivariate data with variable quality,'' in \emph{In
  Proceedings of the IEEE Symposium on Visual Analytics Science and
  Technology}, 2006, pp. 183--190.

\bibitem{Arbesser2017}
C.~Arbesser, F.~Spechtenhauser, T.~M{\"u}hlbacher, and H.~Piringer,
  ``Visplause: Visual data quality assessment of many time series using
  plausibility checks,'' \emph{IEEE Transactions on Visualization and Computer
  Graphics}, vol.~23, no.~1, pp. 641--650, 2017.

\bibitem{Wong2012}
B.~L.~W. Wong and M.~Varga, ``Black holes, keyholes and brown worms: Challenges
  in sense making,'' in \emph{Proceedings of the Human Factors and Ergonomics
  Society 56th Annual Meeting}, 2012, pp. 287--291.

\bibitem{Kirk2014}
A.~Kirk, ``Visualizing zero: How to show something with nothing,''
  http://blogs.hbr.org/2014/05/visualizing-zero-how-to-show-something-with-nothing/,
  May 2014.

\bibitem{Eaton2005}
C.~Eaton, C.~Plaisant, and T.~Drizd, ``Visualizing missing data: graph
  interpretation user study,'' in \emph{Human-Computer Interaction-INTERACT
  2005}.\hskip 1em plus 0.5em minus 0.4em\relax Springer, 2005, pp. 861--872.

\bibitem{Andreasson2014}
R.~Andreasson and M.~Riveiro, ``Effects of visualizing missing data: an
  empirical evaluation,'' in \emph{2014 18th International Conference on
  Information Visualisation}.\hskip 1em plus 0.5em minus 0.4em\relax IEEE,
  2014, pp. 132--138.

\bibitem{Fernandes2018}
M.~Fernandes, L.~Walls, S.~Munson, J.~Hullman, and M.~Kay, ``Uncertainty
  displays using quantile dotplots or cdfs improve transit decision-making,''
  in \emph{Proceedings of the 2018 CHI Conference on Human Factors in Computing
  Systems}, 2018, pp. 1--12.

\bibitem{song2018}
H.~Song and D.~A. Szafir, ``Where's my data? evaluating visualizations with
  missing data,'' \emph{IEEE transactions on visualization and computer
  graphics}, vol.~25, no.~1, pp. 914--924, 2018.

\bibitem{Quinlan1987}
P.~T. Quinlan and G.~W. Humphreys, ``Visual search for targets defined by
  combinations of color, shape, and size: An examination of the task
  constraints on feature and conjunction searches,'' \emph{Perception \&
  psychophysics}, vol.~41, no.~5, pp. 455--472, 1987.

\bibitem{johansson2005}
J.~Johansson, M.~Cooper, and M.~Jern, ``3-dimensional display for clustered
  multi-relational parallel coordinates,'' in \emph{Proceedings IEEE
  International Conference on Information Visualization, IV05}, 2005, pp.
  188--193.

\bibitem{Dayal1994}
B.~S. Dayal, ``Application of feedforward neural networks and partial least
  squares for modelling kappa number in a continuous kamyr digester,''
  \emph{Pulp and Paper Canada}, vol.~95, pp. 26--32, 1994.

\bibitem{UCI}
\BIBentryALTinterwordspacing
K.~Bache and M.~Lichman, ``{UCI} machine learning repository,'' 2013. [Online].
  Available: \url{http://archive.ics.uci.edu/ml}
\BIBentrySTDinterwordspacing

\bibitem{Graziano1993}
A.~M. Graziano and M.~L. Raulin, \emph{Research methods: A process of inquiry
  .}\hskip 1em plus 0.5em minus 0.4em\relax HarperCollins College Publishers,
  1993.

\bibitem{Kahraman2013}
H.~T. Kahraman, S.~Sagiroglu, and I.~Colak, ``Developing intuitive knowledge
  classifier and modeling of users' domain dependent data in web,''
  \emph{Knowledge Based Systems}, vol.~37, pp. 283--295, 2013.

\bibitem{Yeh1998}
I.-C. Yeh, ``Modeling of strength of high performance concrete using artificial
  neural networks,'' \emph{Cement and Concrete Research}, vol.~28, no.~12, pp.
  1797--1808, 1998.

\bibitem{Little2007}
M.~Little, P.~McSharry, S.~Roberts, D.~Costello, and I.~Moroz, ``Exploiting
  nonlinear recurrence and fractal scaling properties for voice disorder
  detection,'' \emph{BioMedical Engineering OnLine}, vol.~6, no.~23, 2007.

\bibitem{Cumming2005}
G.~Cumming and S.~Finch, ``Inference by eye: confidence intervals and how to
  read pictures of data.'' \emph{American psychologist}, vol.~60, no.~2, p.
  170, 2005.

\bibitem{Gorilla}
A.~L. Anwyl-Irvine, J.~Massonni{\'{e}}, A.~Flitton, N.~Kirkham, and J.~K.
  Evershed, ``{Gorilla in our midst: An online behavioral experiment
  builder},'' \emph{Behavior Research Methods}, pp. 1--20, apr 2019.

\bibitem{carsdata}
A.~Asuncion and D.~J. Newman, ``{UCI} machine learning repository. {U}niversity
  of {C}alifornia, {I}rvine, {S}chool of {I}nformation and {C}omputer
  {S}ciences,'' 2007, http://www.ics.uci.edu/$\sim$mlearn/{MLR}epository.html.

\end{thebibliography}
%

\end{balance}

\end{document}